\documentclass[%
 reprint,
groupedaddress,
 amsmath,amssymb,
 aps, pre,
]{revtex4-1}

\usepackage[utf8]{inputenc}
\usepackage[english]{babel}
\usepackage{amsmath}
\usepackage{amsfonts}
\usepackage{amssymb}
\usepackage{makeidx}
\usepackage{graphicx}
\usepackage{arydshln}
\usepackage{cancel}
\usepackage{tcolorbox}
\usepackage{hyperref}
\usepackage{dcolumn}
\usepackage{bm}
\usepackage{colortbl}

\begin{document}
\title{Inference of the Kinetic Ising Model with heterogeneous missing data}
\author{Carlo Campajola}
\email{carlo.campajola@sns.it}
\affiliation{Scuola Normale Superiore di Pisa}
\author{Fabrizio Lillo}
\affiliation{University of Bologna - Department of Mathematics}
\author{Daniele Tantari}
\affiliation{University of Florence - Department of Economics and Management}

\begin{abstract}
    We consider the problem of inferring a causality structure from multiple binary time series by using the Kinetic Ising Model in datasets where a fraction of observations is missing. We take our steps from a recent work on Mean Field methods for the inference of the model with hidden spins and develop a pseudo-Expectation-Maximization algorithm that is able to work even in conditions of severe data sparsity. The methodology relies on the Martin-Siggia-Rose path integral method with second order saddle-point solution to make it possible to calculate the log-likelihood in polynomial time, giving as output a maximum likelihood estimate of the couplings matrix and of the missing observations. We also propose a recursive version of the algorithm, where at every iteration some missing values are substituted by their maximum likelihood estimate, showing that the method can be used together with sparsification schemes like LASSO regularization or decimation. We test the performance of the algorithm on synthetic data and find interesting properties when it comes to the dependency on heterogeneity of the observation frequency of spins and when some of the hypotheses that are necessary to the saddle-point approximation are violated, such as the small couplings limit and the assumption of statistical independence between couplings.
\end{abstract}
\date{\today}
\maketitle
\newpage
\section{Introduction}
Ising-like models and their countless variations have been used throughout the last decades to describe data or model systems with the most diverse nature \cite{bury2013market,Bouchaud2013,tanaka1977model,cocco2017functional,kadirvelu2017inferring} and to increase our understanding of how natural, artificial, social and economic systems work.\\
On the one hand these models, studied in their original physical formulation, can be manipulated to generate a wide range of behaviours mimicking the features of these systems \cite{Bouchaud2013,bornholdt2001expectation}, and use a deductive approach to explain the stylized properties of data we observe in the real world. On the other hand one can use these models in the fashion of descriptive and forecasting models \cite{bury2013market,cocco2017functional,ibuki2013statistical,kadirvelu2017inferring}, by using
Maximum Likelihood (ML) and Maximum A Posteriori (MAP) techniques to fit the model to the data, inductively working towards an explanation of the observations. This is typically regarded to as the inverse formulation of the model, while the former is the direct formulation.\\
A model of this family has recently been revamped for time-series data, the non-equilibrium or Kinetic Ising Model \cite{derrida1987exactly,crisanti1988dynamics}, describing a set of binary units - named ``spins" in the physics literature - that influence each other through time. The simplicity of the model makes it extremely flexible in the kinds of systems it can represent, ranging from networks of neurons in the brain \cite{capone2015inferring} all the way to traders in a financial market \cite{bornholdt2001expectation,SornetteReview}. Recent work on the inverse Kinetic Ising Model has led to the development of exact \cite{sakellariou2013inverse} and Mean Field (MF) \cite{roudi2011dynamical} techniques for the inference of the parameters, and the latter have been used to work with partially observed systems linking to the realm of (Semi-) Restricted Boltzmann Machines \cite{dunn2013learning}. \\
This latest stream of literature sparked our interest for the model applied to time series of financial data at high frequency, where we typically encounter problems related to the lack of homogeneously frequent and synchronized observations \cite{ait2010high,buccheri2017score,Corsi2012}. 
The literature on Kinetic Ising Model has previously considered mainly the inference problem in the presence of hidden nodes \cite{dunn2013learning}, i.e. part of the spins are {\it never} observed, but it is known that they exist and interact with the visible nodes (i.e. spins). This setting is of particular interest in neuroscience where an experiment typically monitors the firing activity of a subset of neurons. In other domains, such as in economics, finance, and social sciences, another type of missing data is often present, namely the case where even for the visible agents (nodes), observations are missing a significant fraction of the times. Moreover in these cases there is a strong heterogeneity of the frequency of observations, i.e. some nodes are frequently observed while other are rarely observed. There are different sources for this lack of data: in some cases, it might be due to the fact the observation is costly for the experimenter, whereas in other cases it is intrinsic to the given problem. Consider, for example, the problem of inferring the opinion of investors from their trading activity. When an investor buys (sells) it is reasonable to assume that she believes the price will increase (decrease), but in many circumstances the investor will not trade leading to missing observations for her belief. Using a suitable inference model, as the one proposed in this paper, it is possible to estimate her belief from the inferred structure of interaction among investors and the observed state of the set of visible ones. We will also include external fields (for example the market price in the previous example) that can influence spins (investors' opinion).

Moving our steps from the work by Dunn et al. \cite{dunn2013learning}, we extend the formulation of the inference procedure to cases where the missing observations are unevenly cross-sectionally distributed, meaning that time series are sampled at a constant rate and whenever no observations are found between two timestamps a missing value is recorded. The result is an algorithm closely related to an Expectation-Maximization (EM) method \cite{expmax1977}, iteratively alternating a step of log-likelihood gradient ascent \cite{Nesterov2008} and the self-consistent resolution of TAP equations \cite{roudi2011dynamical}, that gives as output both a coupling matrix and a maximum-likelihood estimate of the missing values. \\
To evaluate the algorithm performance we devise a series of tests stressing on different characteristics of the input, simulating synthetic datasets with several regimes of intrinsic noise, observation frequency, heterogeneity of variables and model misspecification. We thus define some performance standards that can be expected given the quality of data fed to the method, giving an overview of how flexible the approach is.\\
The paper is organized as follows: in Section 2 we define the considered Kinetic Ising Model, we explain the inference method in detail and describe the approximations needed to make the algorithm converge in feasible time; in Section 3 we present results on synthetic data and give an overview of the performance that can be expected with different data specifications; Section 4 concludes the article.

\section{Solving the Inverse Problem with missing values}
The Kinetic Ising Model (or non-equilibrium Ising Model) \cite{derrida1987exactly} is defined on a set of spins $y \in \lbrace -1, +1 \rbrace^N$, whose dynamics is described by the transition probability mass function
\begin{multline}
    p[y(t+1) \vert y(t) ] = Z^{-1}(t) \exp \Bigg[\sum_{\langle i,j \rangle} y_i (t+1)  J_{ij}y_j(t) +\\
    + \sum_i y_i(t+1) h_i\Bigg]
\end{multline}
where $\langle i,j \rangle$ is a sum over neighbouring pairs on an underlying network, $J_{ij}$ are independent and identically distributed couplings, $h$ is the vector of spin-specific fields and $Z(t)$ is a normalizing constant also known as the partition function.\\
In our treatment of the problem we will adopt a Mean Field (MF) approximation, which relies on the assumption that the dynamics of a spin $i$ depends only on an effective field locally ``sensed" by the spin rather than on the sum of the single specific interactions with others. The result of this picture is that the topology of the underlying network is considered irrelevant and assumed fully connected - although the goal of the inference would be the reconstruction of the network nonetheless - thus the sum on neighbours is substituted by a sum on all the other spins. This recasts the transition probability into the following form
\begin{equation}
p[y(t+1) \vert y(t) ] = Z^{-1}(t) \exp \left[\sum_{i=1}^N y_i (t+1) \tilde{g_i}(t)\right]
\end{equation}
where $\tilde{g_i}(t) = \sum_{j=1}^N J_{ij}y_j(t) + h_i$ is the local effective field of spin $i$ and $J$ is now a square and fully asymmetric matrix with normally distributed entries $J_{ij} \sim \mathcal{N}(0, J_1^2/N)$, where the assumption on the distribution and the scaling of the variance with $N^{-1}$ will be necessary in the forthcoming calculations.\\
Consider observing only a fraction $M(t)/N$ of spins at each time step, and define $G(t)$ as the $M(t) \times N$ matrix mapping the configuration $y(t)$ into the observed vector $s(t) \in \lbrace -1,1 \rbrace^{M(t)}$. Also define $F(t)$ as the $(N-M(t)) \times N$ matrix mapping $y(t)$ into the unobserved spins vector $\sigma(t) \in \lbrace -1,1 \rbrace^{N-M(t)}$. We require that both matrices are right-invertible at all $t$, thus they must have full rank, that implies that observations are not linear combinations of the underlying variables as our interest is in a partially observed system rather than a low-dimensional observation of a high-dimensional system. For the sake of simplicity we assume that the entries are either $0$ or $1$, meaning observation is not noisy or distorted and the right-inverse matrices will coincide with the transpose. \\
In the upcoming calculations we will use some simplifying custom notation in order to reduce what can be some cumbersome equations. We will thus denote $\sideset{}{^\prime}\sum_i$ the sum over indices $i$ at time $t+1$, while the regular $\sideset{}{}\sum_i$ indicates a sum over indices $i$ at time $t$ and $\sideset{}{^-}\sum_i$ a sum at time $t-1$. Accordingly, we will indicate with $s_i$ spin $i$ at time $t$, with $s^-_i$ at time $t-1$ and with $s_i^\prime$ at time $t+1$, and the same applies for $g$, $\sigma$ and any other variable. Also indices $i,j,k,l$ are used for observed variables, whereas indices $a,b,c,d$ will identify unobserved variables.\\
In this notation, the probability mass function is rewritten as
\begin{align}
p[\lbrace s^\prime, \sigma^\prime \rbrace \vert \lbrace s, \sigma \rbrace ] = Z^{-1}  \exp \left[ \sideset{}{^\prime}\sum_{i} s_i^\prime g_i^\prime + \sideset{}{^\prime}\sum_{a} \sigma_a^\prime g_a^\prime \right] \label{eq:pmf}
\end{align}
Defining the matrices $J^{oo}(t+1) = G(t+1)JG^T(t)$, $J^{oh}(t+1) = G(t+1)JF^T(t)$, $J^{ho}(t+1) = F(t+1)JG^T(t)$ and $J^{hh}(t+1) = F(t+1)JF^T(t)$ the local fields are
\begin{align}
g_i = \sum_{j} J_{ij}^{oo} s_j^- + \sum_{b}J^{oh}_{ib} \sigma_b^- +h_i \nonumber \\
g_a = \sum_{j} J_{aj}^{ho} s_j^- + \sum_{b}J^{hh}_{ab} \sigma_b^- +h_a \label{eq:fielddef}
\end{align}
and the partition function or normalization constant is
\begin{equation*}
    Z = \sideset{}{^\prime}\prod_{i,a} 2 \cosh (g_i^\prime) 2 \cosh (g_a^\prime)
\end{equation*}
The ultimate purpose of this work is to devise a method to obtain Maximum Likelihood Estimates (MLE) for the parameters $J,h$ and the unobserved spins $\sigma$. The likelihood function is just the product through time of the independent transition probabilities expressed in Eq. \ref{eq:pmf}, taking the trace over the missing values
\begin{equation}
p[\lbrace s \rbrace ] = \mathrm{Tr}_{\sigma} \prod_t p[\lbrace s^\prime, \sigma^\prime \rbrace \vert \lbrace s, \sigma \rbrace ] \label{eq:likelihood}
\end{equation}
To solve the problem, our approach is closely related to the one developed by Dunn et al. \cite{dunn2013learning}, where the authors investigate on a system where only a subset of spins is observable. The extension to our case is presented below.

The trace of Eq. \ref{eq:likelihood} is non-trivial to be done. However the Martin-Siggia-Rose path integral formulation \cite{msr1973} allows to decouple spins and perform the trace at the cost of computing a high dimensional integral. Define the functional
\begin{equation}
\mathcal{L}[\psi] = \log \mathrm{Tr}_{\sigma} \prod_t \exp \left[ \sum_a \psi_a \sigma_a \right] p[\lbrace s^\prime, \sigma^\prime \rbrace \vert \lbrace s, \sigma \rbrace ]
\end{equation}
Notice that this is equivalent to the log-likelihood if $\psi_a(t) = 0$ $\forall a,t$, thus the goal of the calculation will be to efficiently maximise $\mathcal{L}[\psi]$ in the $J, h$ coordinates considering the limit when $\psi \rightarrow 0$. As will become clear in the next steps, the introduction of these so-called ``auxiliary fields" is necessary to switch from the unknown values $\sigma$ to their posterior expectations $m$, thus smoothing the log-likelihood function eliminating unknown binary variables from its formula.
Call 
\begin{align*}
Q[s, \sigma] = \sum_t \sum_{i}&s_i g_i + \sum_t \sum_{a} \sigma_a g_a + \\ 
- \sum_t \sum_{i}&\log 2\cosh(g_i) - \sum_t \sum_{a} \log 2 \cosh(g_a) \\
\Delta = \sum_t \sum_{i}& i \hat{g}_i \left[ g_i - \sum_{j} J^{oo}_{ij}s_j^- - \sum_{b} J^{oh}_{ib} \sigma_b^- - h_i \right] + \\
+ \sum_t \sum_{a}& i \hat{g}_a \left[ g_a - \sum_{j} J^{ho}_{aj} s_j^- - \sum_{b} J^{hh}_{ab} \sigma_b^- - h_a \right]
\end{align*}
where $e^\Delta$, integrated over the $\hat{g}$s is the integral representation of the Dirac delta function. Then one obtains
\begin{equation}\label{lagrangian}
\mathcal{L}[\psi] = \log \int \mathcal{DG} \exp [\Phi]
\end{equation}
where $\mathcal{G} = \lbrace g_i, g_a, \hat{g}_i, \hat{g}_a \rbrace_t$ and
\begin{equation}
\Phi = \log \mathrm{Tr}_{\sigma} \exp \left[Q + \Delta + \sum_t \sum_{a} \psi_a \sigma_a \right]
\end{equation}
Now the trace can be easily computed since the introduction of the delta function has decoupled the $\sigma$s by fixing the value of the local fields $g$.

As mentioned, the cost is computing the integral of Eq. \ref{lagrangian}, which can be solved via the saddle-point approximation, where the saddle-point is obtained by the extremization of $\Phi$ with respect to the coordinates in $\mathcal{G}$.\\
The missing part of the puzzle is the posterior mean $\mathbb{E}\left[\sigma_a (t) \right]$, for which $\mathcal{L}$ acts as the generating functional
\begin{equation*}\mathbb{E}\left[\sigma_a (t)\right]= m_a(t) = \lim_{\psi_a(t) \rightarrow 0} \mu_a(t) = \lim_{\psi_a(t) \rightarrow 0}\frac{\partial \mathcal{L}}{\partial \psi_a(t)}
\end{equation*}
where the expectation is performed under the posterior measure $p[\lbrace \sigma \rbrace \vert \lbrace s, J, h \rbrace]$. \\

This zero-order approximation is rather rough, nonetheless the saddle-point method can be solved at higher orders of approximation.\\
The second-order (\textit{i.e.} Gaussian) correction to the saddle point solution of the integral in Eq. \ref{lagrangian} is 
\begin{equation*}
    \delta \mathcal{L} = - \frac{1}{2}\log \det [\nabla^2_{\mathcal{G}} \mathcal{L}]
\end{equation*}
where $\nabla^2_{\mathcal{G}} \mathcal{L}$ is the Hessian matrix in the $\mathcal{G}$ space of $\mathcal{L}$ evaluated at the saddle point. The resulting structure of the matrix, shown in the Supplementary Material for the sake of space, is sparse and almost block-diagonal.

We are interested in the determinant, and in particular its logarithm. Dividing the Hessian in the matrices $\alpha$ containing block-diagonal elements and $\beta$ containing the rest, we find
\begin{multline}
\log\det(\alpha+\beta)=\log\det(\alpha) + \log\det[\mathbb{I}+\alpha^{-1}\beta] =\\
= \log\det(\alpha) + \mathrm{Tr}\log[\mathbb{I} + \alpha^{-1}\beta] \approx \\
\approx \log\det(\alpha) + \mathrm{Tr}[\alpha^{-1}\beta] + \frac{1}{2} \mathrm{Tr}\lbrace [\alpha^{-1} \beta]^2 \rbrace + ... \label{eq:approx}
\end{multline}
Given that $\alpha$ is block-diagonal, so will be $\alpha^{-1}$, then $\mathrm{Tr}[\alpha^{-1}\beta]=0$ and we ignore higher order terms assuming the off-diagonal part of the Hessian matrix is small compared to the diagonal one. In our initial assumption, the couplings $J_{ij}$ are Gaussian random variables with mean of order $1/N$ and variance of order $J^2_1/N$, which means $\log \det (\alpha)$ is quadratic in $J_1$ (see Supplementary Material). The determinant now can be computed and a weak couplings expansion (i.e. $J_1 \rightarrow 0$) can be made to eliminate the logarithm, leading to the final approximate form of the correction
\begin{align*}
\delta \mathcal{L} \approx & -\frac{1}{2} \sum_t \sideset{}{'}\sum_{i} \left[ \left(1-\tanh^2(g_i^\prime) \right) \sum_{b} \left[J^{oh \prime}_{ib}\right]^2(1-\mu_b^2)\right] + \\ 
&- \frac{1}{2} \sum_t \sideset{}{'}\sum_{a} \left[ \left(\mu_a^{\prime \, 2} - \tanh^2(g_a^\prime) \right) \sum_{b} \left[ J^{hh \prime}_{ab} \right]^2 (1-\mu_b^2)\right] 
\end{align*}

Given the new form of $\mathcal{L}_1 = \mathcal{L}_0 + \delta \mathcal{L}$, we need to recalculate the self-consistency relation for $m_a(t)$ and the learning rule for $J$. As for $m_a(t)$, we can easily see that it is going to coincide with $m_a(t) = \lim_{\psi_a(t) \rightarrow 0} \mu_a(t) + l_a(t)$, where 

\begin{equation}
l_a(t) = \frac{\partial( \delta \mathcal{L})}{ \partial \psi_a(t)}
\label{eq:corrselfcon}
\end{equation}

Implementing the MSR method has introduced an explicit dependence of the $\mathcal{L}$ functional from the auxiliary fields $\hat{g}$ and $\psi$, which however make little sense in terms of the model itself. Now that we have solved the integral at the saddle-point and in its immediate neighbourhood the auxiliary fields can be absorbed back into the original variables by performing a Legendre transform of $\mathcal{L}$, exploiting the fact that $\mathcal{L}$ is convex and that we would rather have it depend on the conjugate field of $\psi$, that is $\mu$. The transform is
\begin{equation}
\Gamma [\mu] = \mathcal{L} - \sum_t \sum_{a} \psi_a(t) \mu_a(t) \; s.t. \; -\psi_a(t) = \frac{\partial \Gamma[\mu]}{\partial \mu_a(t)}
\end{equation}

and so we can adopt $\Gamma$ as the functional to be maximised in the learning process instead. At zero-order, this is easily found to be
\begin{multline}
\Gamma_0 [\mu] = \sum_t \Bigg[ \sideset{}{'}\sum_{i} \left[ s_i^\prime g_i^{0 \, \prime} - \log 2\cosh (g_i^{0 \, \prime}) \right] + \\ + \sideset{}{'}\sum_{a} \left[ \mu_a^\prime g_a^{0 \, \prime} - \log 2 \cosh(g_a^{0 \, \prime}) \right] + \sum_{a} S[\mu_a] \Bigg]
\end{multline}

where $S[x] = -\frac{1+x}{2} \log (\frac{1+x}{2}) - \frac{1-x}{2} \log(\frac{1-x}{2})$ is the entropy of an uncoupled spin with magnetization $x$. It is relevant to mention that so far the functional is expressed in terms of $\mu$, while we have already highlighted that after the Gaussian correction a new term $l$ is introduced in the formula for $m$. However, since we are restricting to second order in $J$, the terms containing $l$ in $\Gamma$ are all of superior order and are thus negligible in this approximation, then $\Gamma_0[m] \approx \Gamma_0[\mu]\vert_{\mu = m}$. Performing the exact same steps on the correction term $\delta \mathcal{L}$ one finds the corrected functional
\begin{equation*}
    \Gamma_1[m] = \Gamma_0 [m] + \delta \mathcal{L}[m]
\end{equation*}
$\Gamma_1$ is the functional to be optimized through an Expectation-Maximization-like algorithm, recursively computing the self-consistent magnetizations $m$ given $J,h$ and then climbing the gradient $\nabla_{J,h} \Gamma_1$ to obtain a new $J$ matrix and $h$ vector. \\
Once the log-likelihood is maximized and the final iteration of the expectation part of the algorithm is finished, the result is a Maximum Likelihood Estimate of the couplings as well as a Maximum A Posteriori estimate of the hidden spins $\sigma$, given by $\hat{\sigma} (t) = \mathrm{sign} (m_t)$.\\

Summarizing, the procedure is the following:
\begin{tcolorbox}[title=\textbf{Algorithm},colbacktitle=white,coltitle=black]
\begin{itemize}
\item Initialize $J$, $h$, $m(t)$
\item Until convergence is reached
\begin{itemize}
\item compute the self-consistent magnetizations $m(t)$
\item compute the gradient $\nabla_{J,h} \Gamma_1$
\item apply Gradient Ascent step, in our case Nesterov's II method proximal gradient ascent with backtracking line search
\end{itemize}
\item Possibly involve LASSO $\ell_1$-norm regularization or pruning techniques to obtain a sparse model.
\end{itemize}
\end{tcolorbox}

\section{Tests on synthetic data}

We perform a series of tests on the algorithm in order to assess its performance in several diverse conditions of data availability. We particularly focus on how we select the observed spins and on the structure of the coupling matrix $J$ in the data generating model. To construct the $G(t)$ and $F(t)$ matrices, we assign to each spin a probability $p_i$ of being observed, meaning that $y_i(t)$ is observed with probability $p_i$ for all $t$.\\
We explore how the performance of the inference depends on the following model specifications:
\begin{enumerate}
\item[0.] The average observation frequency, taking the Bernoulli probabilities $p_i = p$, $\forall i=1,\ldots, N$;
\item The heterogeneity of the Bernoulli probabilities $p_i$, which we choose to be distributed according to a Beta distribution $B(a(K), b(K))$ with given mean $K$ and shape parameters $a$ and $b$;
\item The scale $J_1$ of the $J$ entries, which are distributed as $J_{ij} \sim \mathcal{N}(0,J_1^2/N)$; 
\item The structure of the $J$ matrix, specifically whether the underlying network is fully connected or an Erd\H{o}s-R\'{e}nyi random network of varying density, adopting either the LASSO $\ell_1$ regularization \cite{tibshirani1996regression} or the decimation procedure \cite{decelle2015inference} to select the links;
\item The asymmetry of the $J$ matrix. One of the key assumptions in the calculation is that $J_{ij} \neq J_{ji}$ and that they are independent and identically distributed, and we investigate how far one can violate it up to the case of a symmetric $J$ matrix;
\item The dependency on the length of the time series relative to the number of units involved, $T/N$, to check the estimate asymptotic efficiency.
\end{enumerate}
In Test 0 we study the performance of the algorithm in a very simple setting of missing information, where each variable has the same probability of being observed and the generating model is a fully-connected Kinetic Ising model. This is intended to study the effect the average amount of missing information in the sample has on the inference, without considering the possibility of having heterogeneous types of nodes. In this setting we also introduce a procedure we call Recursive E-M: by properly iterating the algorithm multiple times  it allows to boost data artificially  thus achieving good performances even when the fraction of missing values is particularly high. \\
In Test 1 we explore the possibility that spins have heterogeneous observational properties. We sample the $\lbrace p_i\rbrace$ from a Beta distribution varying parameters to probe different levels of heterogeneity. The Beta distribution allows to range from a sharply peaked unimodal distribution to a sharply peaked bimodal distribution tuning the shape parameters $\alpha$ and $\beta$, while keeping the mean $K$ constant: the former case is a situation of perfect homogeneity in the frequency of observations calling back to Test 0, while the latter is the extreme heterogeneity of having some units that are (almost) always hidden while the others are (almost) always observed. We select some intermediate cases to characterize how heterogeneity in observation frequency affects the identification of the model parameters. \\
Test 2 aims at assessing whether there is a minimal interaction strength to have the inferential process converging and how the approximations necessary to develop the method impact the accuracy of the inference. Indeed while $J_1$ in the physical model is proportional to the ratio between the strength of the magnetic coupling interaction and the temperature at which the system is observed, from a modelling perspective it is inversely proportional to the impact of the noise on the dynamics. Given the approximation of Eq. \ref{eq:approx}, if $J_1$ gets too large, the precision with which the parameters are identified should get worse. We thus expect to find an optimal region for the inference to be accurate, bounded from below by an identifiability threshold and from above by the limit of validity of the expansion.\\
In Test 3 we pursue the goal of making the methodology useful for real world scenarios, where it is highly unlikely that all spins interact among themselves and the underlying network is probably sparse. We compare the performance of two well established techniques, the LASSO $\ell_1$ regularization and the decimation procedure, and explore how these two methods perform paired with our algorithm by simulating data on a set of Erd\H{o}s-R\'{e}nyi random networks with different densities. \\
In a similar spirit, in Test 4 we study how the i.i.d. assumption made in Eq. \ref{eq:approx} affects the performance in situations where coupling coefficients are pairwise correlated or even symmetric, a condition we envision to be more realistic in social and economic environments \cite{squartini2013reciprocity}. We vary the correlation parameter $\mathrm{Cor}(J_{ij}, J_{ji})=\rho$ for $i\neq j$ between 0 and 1, with the symmetric case being also of special interest because the model transforms into a dynamical form of the Sherrington-Kirkpatrick model, thus connecting to the extensive literature on the topic.\\
Finally, a sanity check is made in Test 5 by looking at the dependency of performance metrics on the ratio $T/N$, that is the ratio between the number of observations and the number of spins, to characterize the convergence rate of the estimator towards the true value and its consistency. \\
We will test the algorithm and evaluate the performance using mainly two metrics, one relative to the reconstruction of the couplings and one to the reconstruction of missing values:
\begin{enumerate}
\item The Root Mean Square Error (RMSE) on the elements of the matrix $J$, $\mathrm{RMSE} = \sqrt{\langle (\hat{J}_{ij} - J_{ij})^2 \rangle_{ij}}$, suitably rescaled when comparing experiments with different $J_1$;
\item The ``Reconstruction Efficiency" (RE), namely the fraction of spins that are correctly guessed among the hidden ones averaged throughout the time series, or $\mathrm{RE} = \langle \frac{1}{N-M(t)}\sum_a\delta_{\hat{\sigma}_a (t), \sigma_a (t)} \rangle_t $
\end{enumerate}

\subsection{Test 0: dependency on a homogeneous $p_i$}

\begin{figure}[h]
\includegraphics[width=1\linewidth]{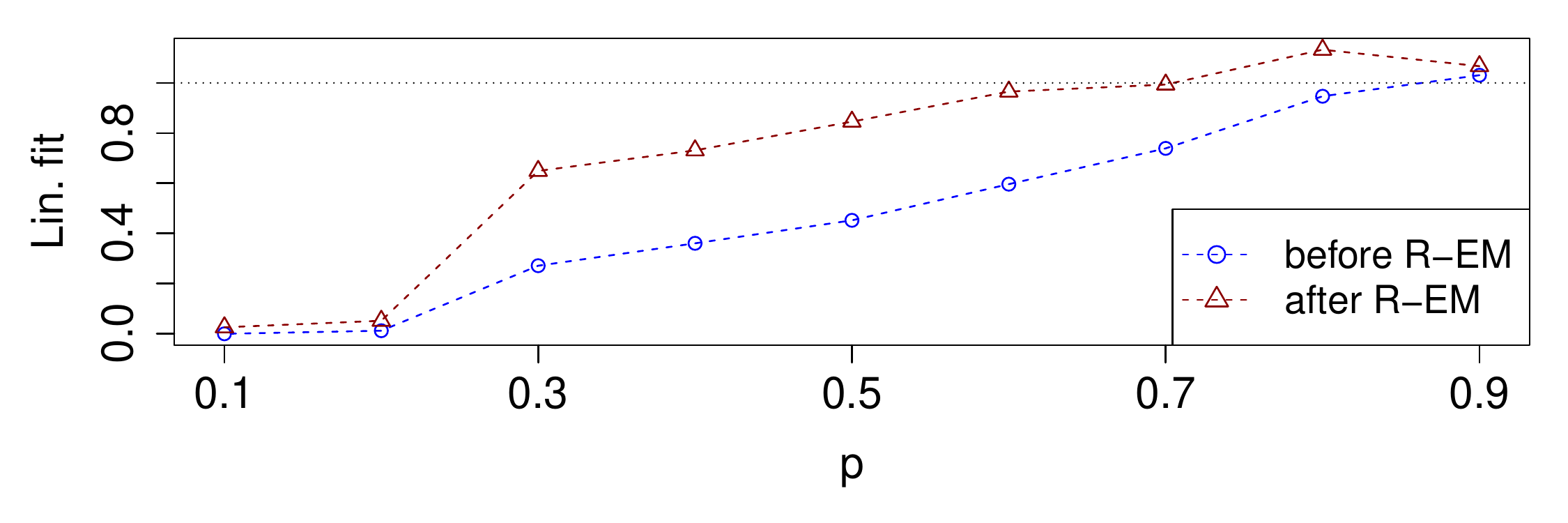}
\includegraphics[width=1\linewidth]{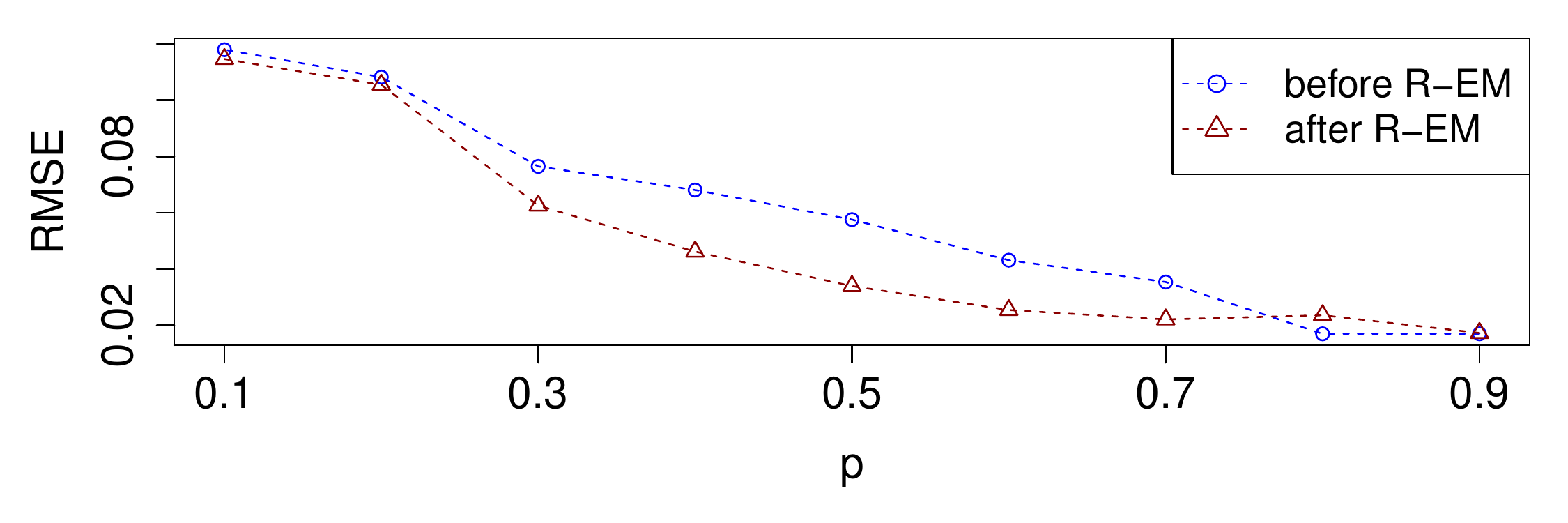}
\includegraphics[width=1\linewidth]{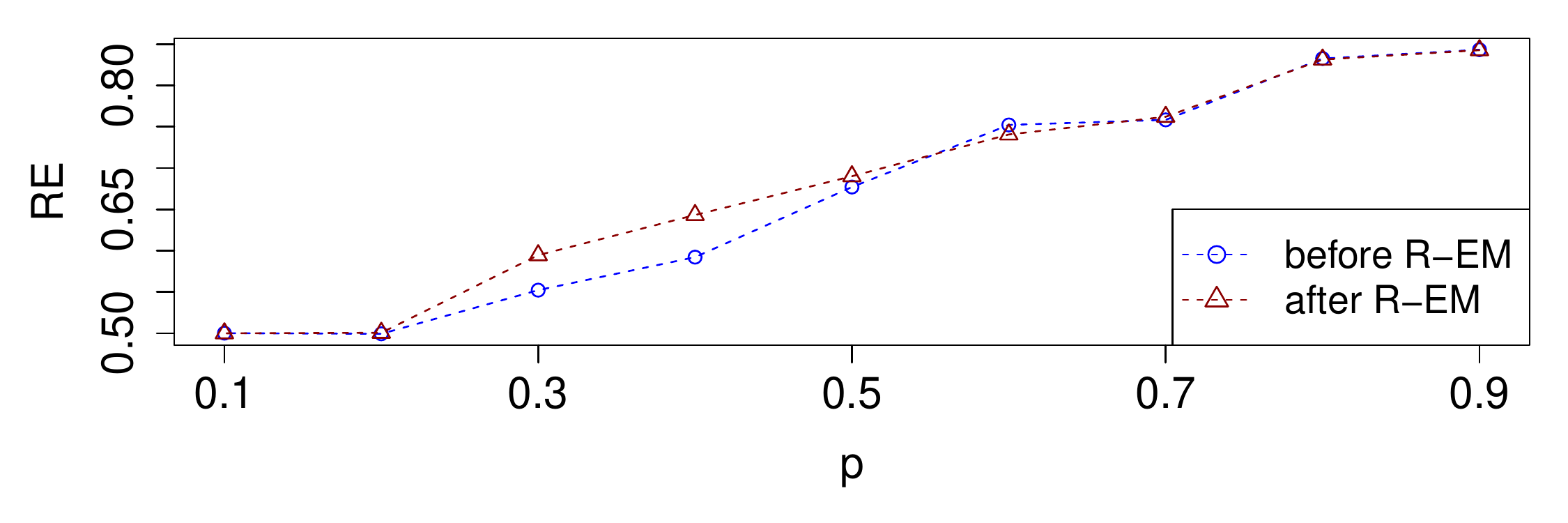}
\caption{(top) Angular coefficient of the linear fit $\hat{J}_{ij} = a J_{ij} + c$ before and after R-EM varying the average observation density $p$; (middle) Root Mean Squared Error on the couplings; (bottom) Reconstruction Efficiency.}\label{fig:test0}
\end{figure}

The algorithm is outstandingly resilient to cases with few observations available. We simulate a system of $N=100$ spins, for $T=10000$ time steps, with $J_{ij} \overset{iid}{\sim} \mathcal{N}(0,1/N)$ lying on a fully connected network and we give a probability of observation to each variable $p_i=p$, with $p$ ranging from $0.1$ to $0.9$. As can be seen from the top panel of Figure \ref{fig:test0}, showing the linear regression coefficient $a$ of $\hat{J}_{ij} = a J_{ij} + c$, with one iteration of the method we get a very reliable result for the couplings for $p \geq 0.8$, although below this value the lack of data reduces the quality of the estimation and moves the estimates towards $0$. To overcome this issue, we propose the aforementioned R-EM procedure as a further enhancement of our algorithm: once a maximum of the likelihood has been reached, a fraction of hidden spins is substituted with their maximum likelihood estimates $\hat{\sigma}_a = \mathrm{sign} (m_a)$ and the inference is run again on the new, artificially boosted data. Since $m$ is proportional to the probability of the spin being up, we choose the missing values to be substituted at every $t$ as the ones with the most polarized magnetization, \textit{i.e.} for which $m$ is closer to $\pm 1$. This artificial boosting on the data shows promising results since with a few recursions the performance is noticeably better even in cases with severe lack of observations, as is also reflected in the middle and bottom panels of Figure \ref{fig:test0}. We defer a more rigorous treatment of this recursive method to future work, while still proposing it here as we find it surprisingly accurate.\\
The bottom panel of Figure \ref{fig:test0} shows the Reconstruction Efficiency, which gets worse almost linearly as the number of observations decreases and on which the R-EM has a smaller effect, albeit still being a clear improvement. It is evident from all panels that when a large fraction of data is missing ($p \leq 0.2$) the inference fails to identify any of the parameters and the model is no better than a coin flip at reconstructing configurations.

\subsection{Test 1: heterogeneous $p_i$}

In Test 1 we want to highlight how our model is a generalization of the one studied extensively by Dunn et al. \cite{dunn2013learning} and to characterize the impact of heterogeneity on the inference performance. To give a better comparison with the aforementioned paper, we realize simulations morphing from our initial specification of $p_i = p \; \forall i$, studied in Test 0, to a case very close to the one of Dunn et al. where $p_i \in \lbrace 0, 1 \rbrace$, that is some variables are always observed and some are always hidden. We choose to take the probabilities distributed according to a Beta distribution, $p_i \sim B(a(K),b(K))$, giving us the possibility of leaving the average number of observations constant while skewing the distribution between a fully bimodal (small $b(K)$) and a sharp quasi-delta function (large $b(K)$). We choose the parameters $a$ and $b$ such that the mean $\mathbb{E}[p_i] = K$ is constant, so that different tests can be compared and the role of heterogeneity is highlighted. This binds the values of $a$ and $b$ through $a = \frac{K b}{1-K}$.

\begin{figure}
\includegraphics[width=.49\linewidth]{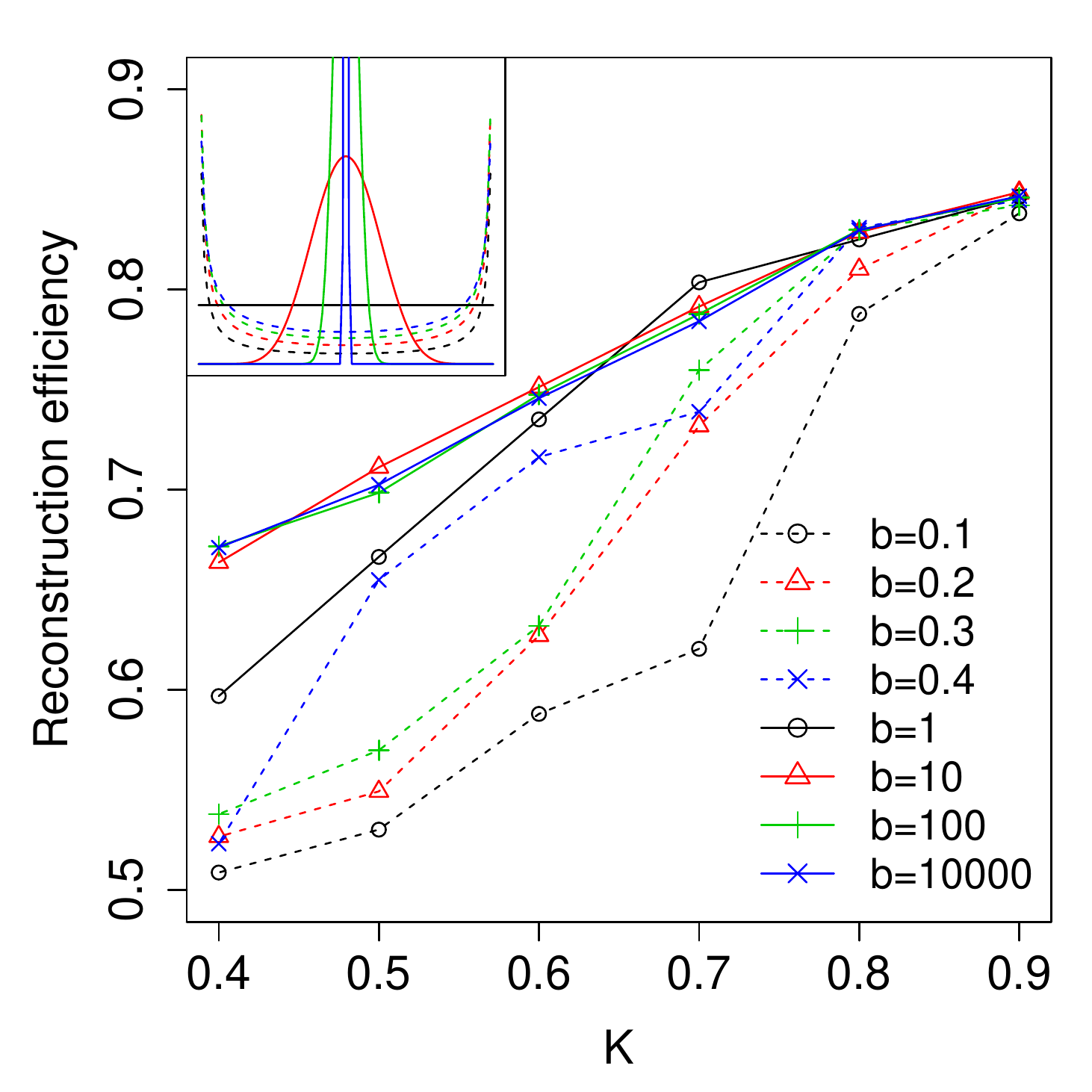}
\includegraphics[width=.49\linewidth]{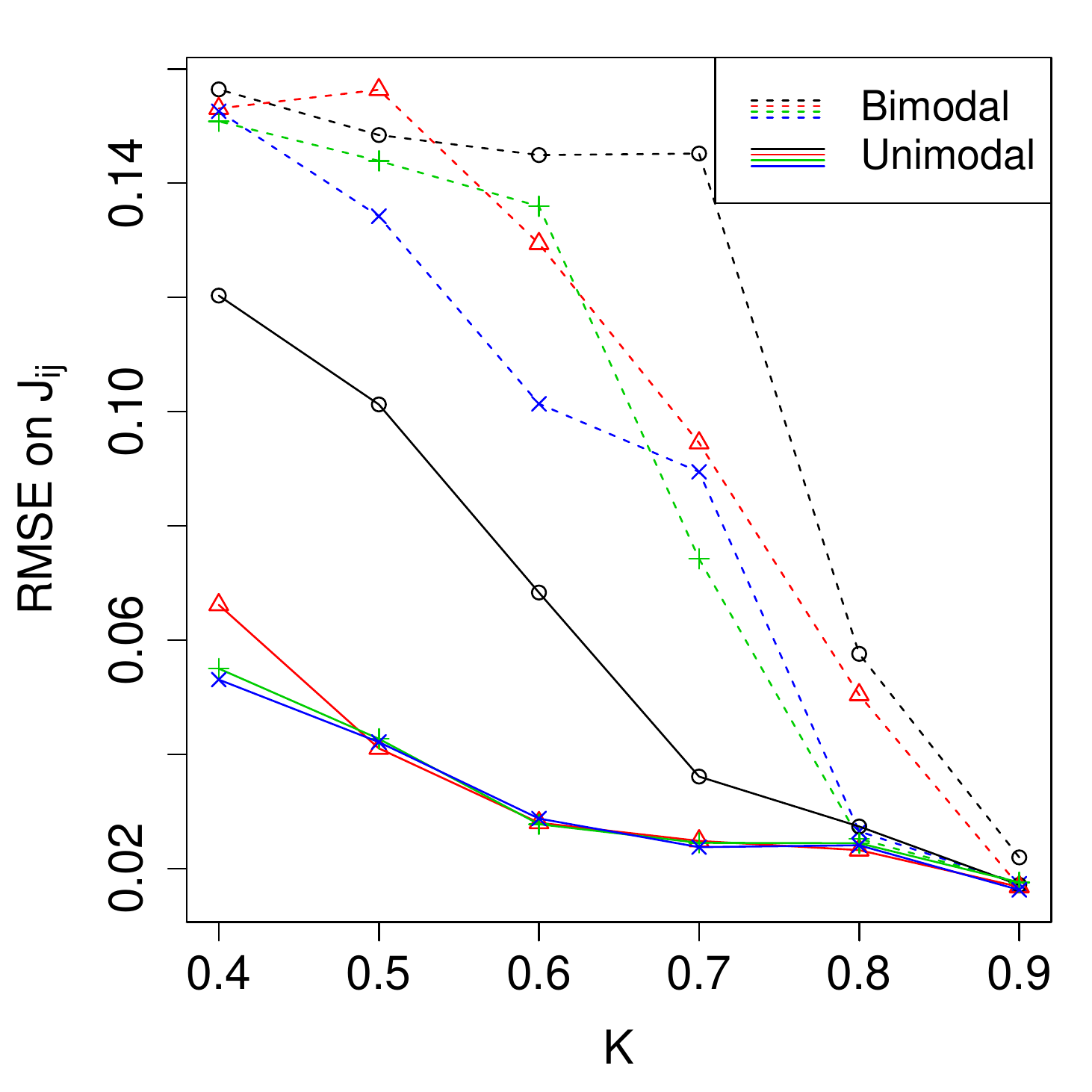}
\caption{(left) Reconstruction efficiency as a function of $K$ with different Beta parameters. Inset: the pdf of the adopted Beta distributions with $K=0.5$ (color coding is the same as in the main panel) (right) Root Mean Square Error on the couplings as a function of $K$ with different Beta parameters.}\label{betadep}
\end{figure}

The results of Figure \ref{betadep} clearly show that when the distribution is bimodal, that is when some variables are very rarely observed, the performance of the algorithm is worse. With a sample size of $T=10^4$ and $N=40$, the Dunn et al. model approximated by $B(a(K), 0.1)$ is identified with reasonable performance only when $K\geq 0.8$. This is extremely mitigated when the observations are more homogeneously distributed, particularly in the case of the coupling coefficients whose estimation seem to require a rather homogeneous distribution of observations among variables to be reliable. On the other hand, the reconstruction efficiency is far less demanding in terms of data quality and a reasonable performance is achieved even with sparse data and heterogeneous observations.

\begin{figure}
    \centering
    \includegraphics[width=1\linewidth]{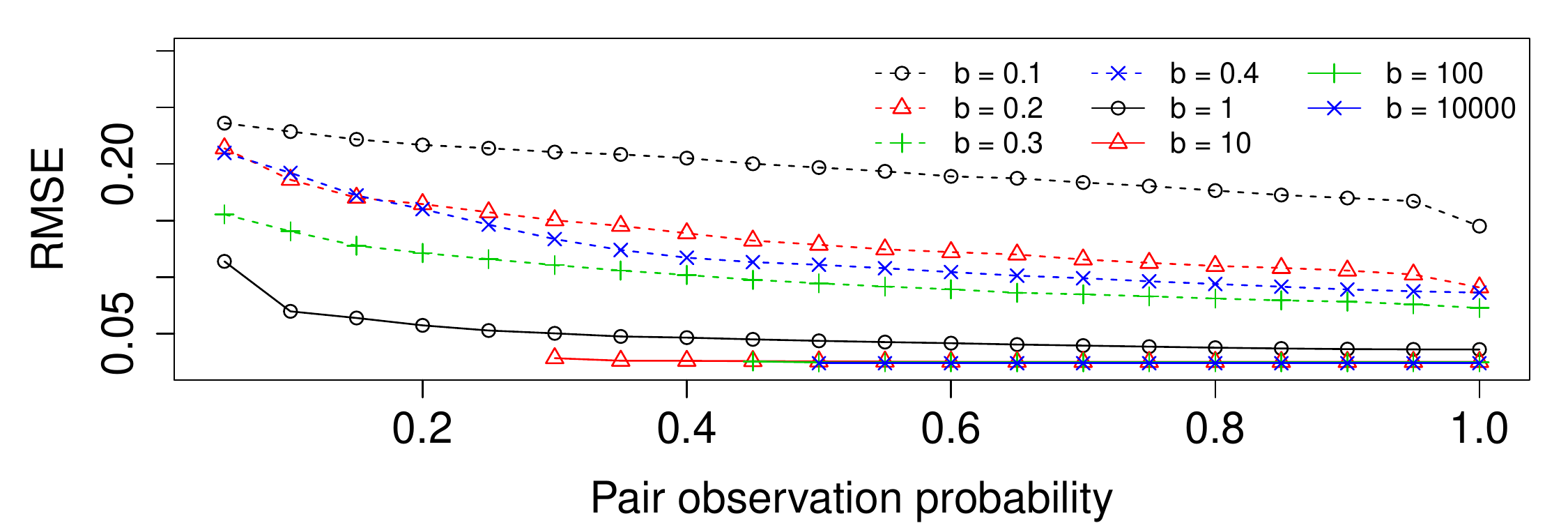}
    \includegraphics[width=1\linewidth]{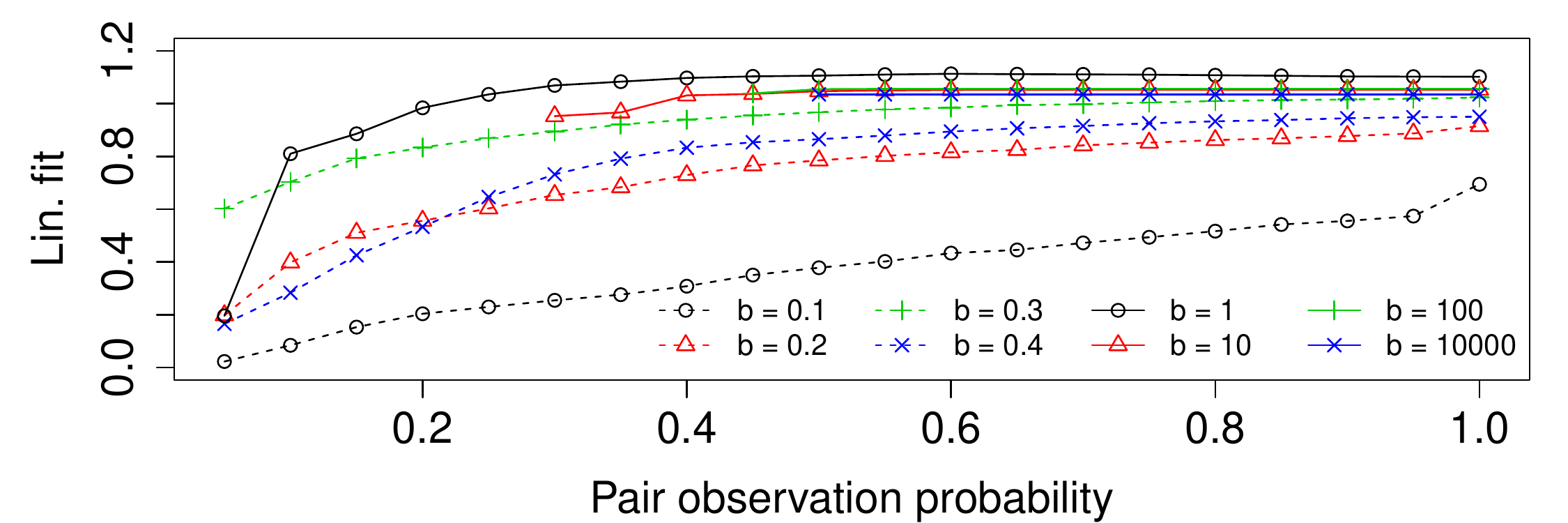}
    \caption{Quality of inference varying the probability of observing the end nodes at subsequent times. (top) RMSE for different values of the Beta $b$ parameter with mean $K=0.7$; (bottom) Linear fit coefficient for different values of the $b$ parameter, $K=0.7$.}
    \label{fig:test1b}
\end{figure}

In Figure \ref{fig:test1b} we  plot the Root Mean Square Error on couplings conditional on the probability of observing subsequently the spins at their ends. This probability is simply given by $p_{ij} = p_i p_j$ since observations are independently sampled, and the RMSE is
\begin{equation*}
    \mathrm{RMSE}(p) = \sqrt{\langle (\hat{J}_{ij} - J_{ij})^2 \rangle_{p_{ij} = p}}
\end{equation*}

where the mean is taken on links that have (close to) the same joint observation probability. The plots highlight how the least observed the pair, the worse the precision of the fit, however it is also clear that the error grows for the more frequently observed couplings too. This is partially mitigated when one looks at the linear fit between the inferred $J$s and the true ones, meaning that the error is mostly affected by the variance component rather than the bias one.\\
The overall effect of heterogeneity is thus a decrease in the quality of the inference, with a stronger effect on couplings that are between the least observed pairs of spins and an important loss in accuracy, but with a bias component that is mitigated for the most frequently observed pairs.

\subsection{Test 2: dependency on $J_1$}

\begin{figure}[h]
\includegraphics[width=1\linewidth]{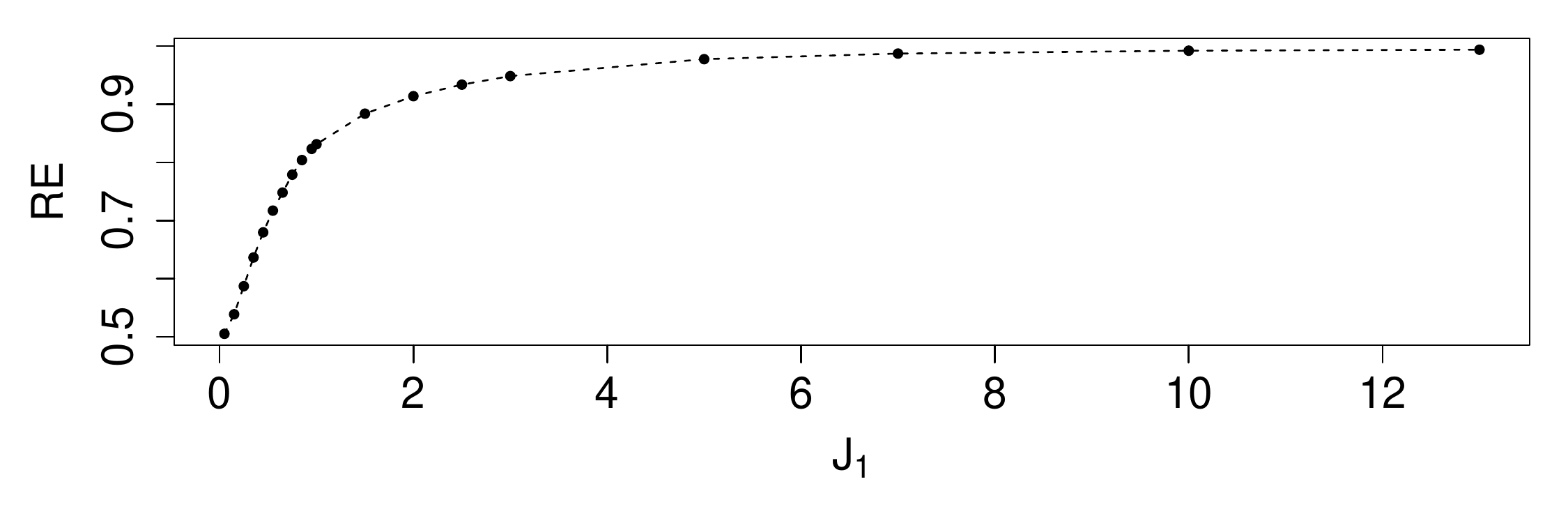}
\includegraphics[width=1\linewidth]{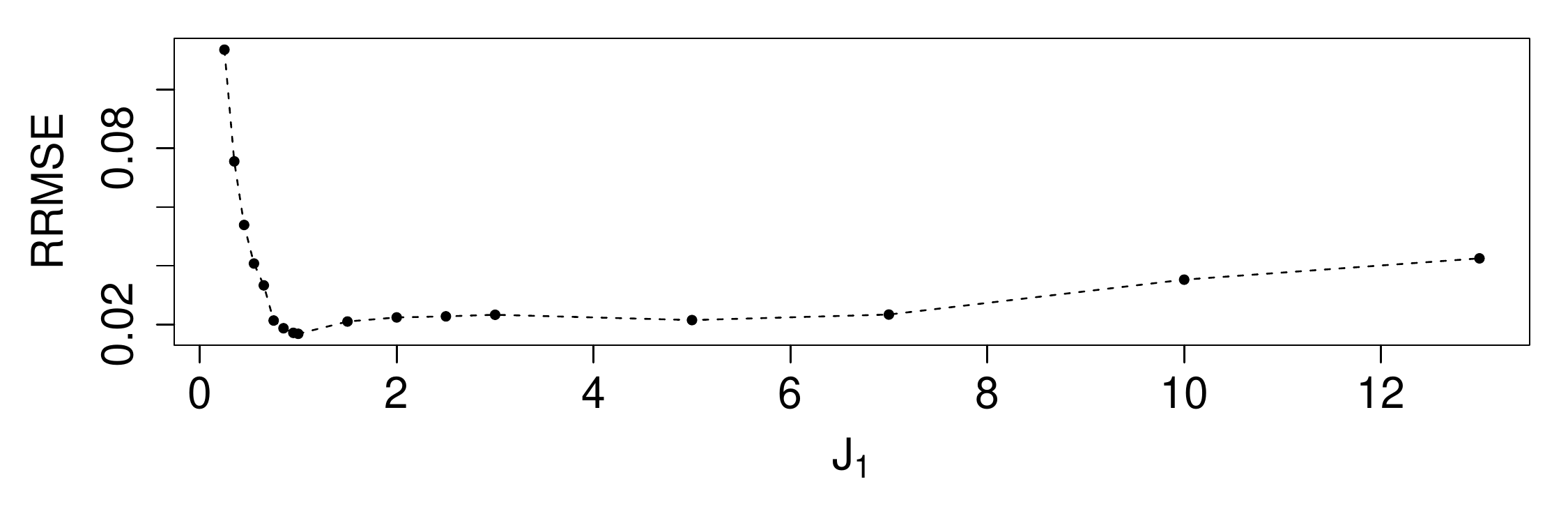}
\caption{(top) Reconstruction Efficiency as a function of $J_1$. (bottom) Rescaled RMSE (by $J_1$) on the couplings as a function of $J_1$.}\label{fig:jdep}
\end{figure}

So far we have dealt with elements of $J$ drawn i.i.d. from a $\mathcal{N}(0,1/N)$ distribution. We want to relax this hypothesis and, while changing the mean value of the distribution would not be particularly meaningful in that it would just shift the correlation patterns between variables, it makes sense to investigate the behaviour as one changes the variance and thus the strength of the interactions. While there is no phase transition in the underlying model as long as the $J_{ij}$ are i.i.d., we want to check how weak can the couplings be in order to be correctly inferred and give a reliable reconstruction of the data. In other words, we are trying to identify a threshold in the interaction strength below which the algorithm is unable to converge. \\
We report results for an experiment with $N=100$, $T=10000$, $p_i=p=0.8$ and $J_1$ ranging from $0.05$ to $13$. We see from Figure \ref{fig:jdep} that increasing the typical size of couplings positively affects the quality of the inference, as should be expected since the dynamics is less affected by randomness. In the top panel we plot the reconstruction efficiency which has a steady increase and saturates towards $1$ after $J_1 \simeq 5$. The bottom panel shows the relative RMSE, that is $\mathrm{RMSE}/J_1$, and we see that it drops below $5\%$ for $J_1 > 0.5$. It is rather surprising to see how, regardless of the small couplings expansion we utilize in Eq. \ref{eq:approx}, the algorithm seems to work efficiently even in cases where the variance of the couplings $J_1^2/N$ is of order $1$, albeit a region of optimality for the inference of the couplings seems to lie within $0.5 \leq J_1 \leq 7$.

\subsection{Test 3: impact of network structure}

\begin{figure}
\centering
\includegraphics[width=.49\linewidth]{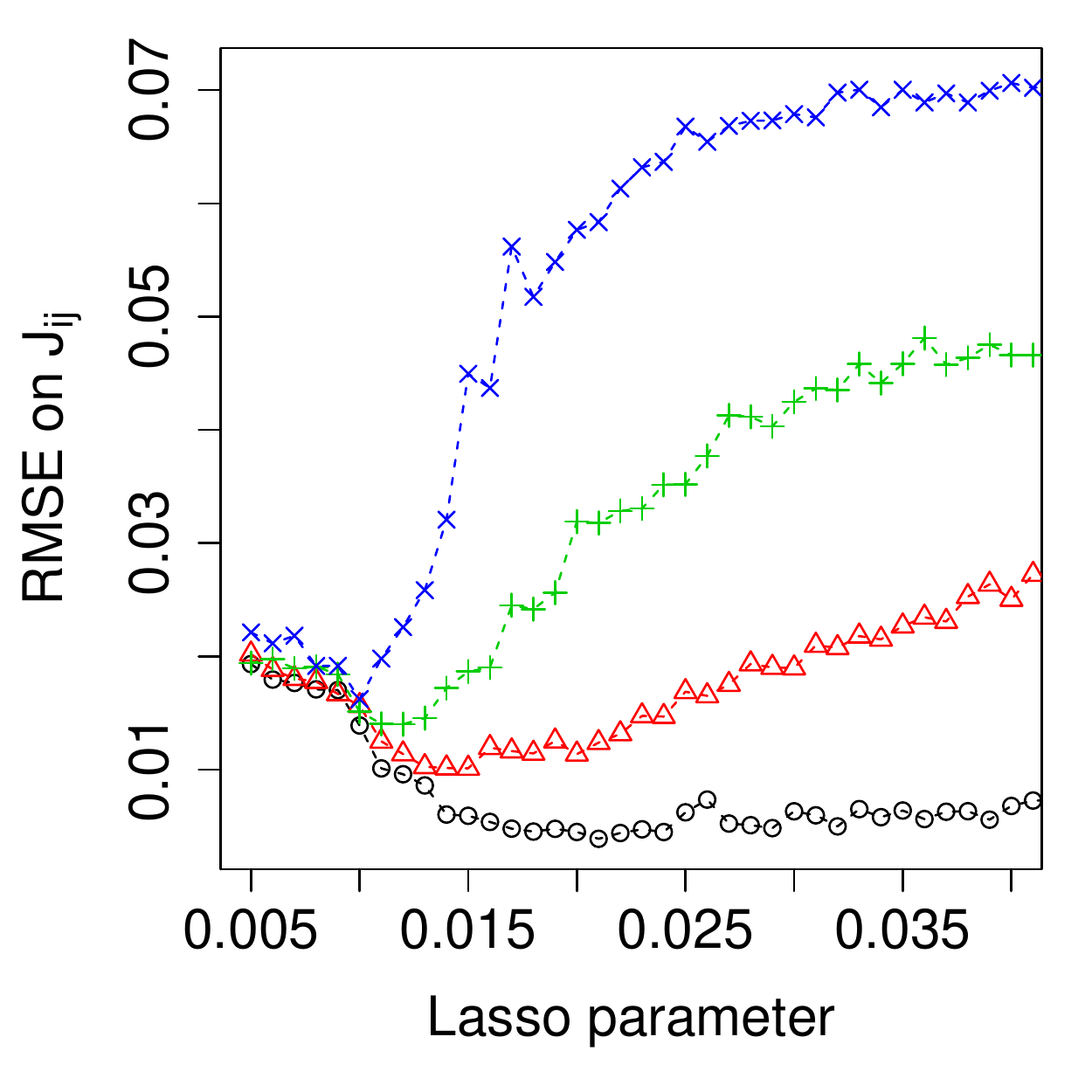}
\includegraphics[width=.49\linewidth]{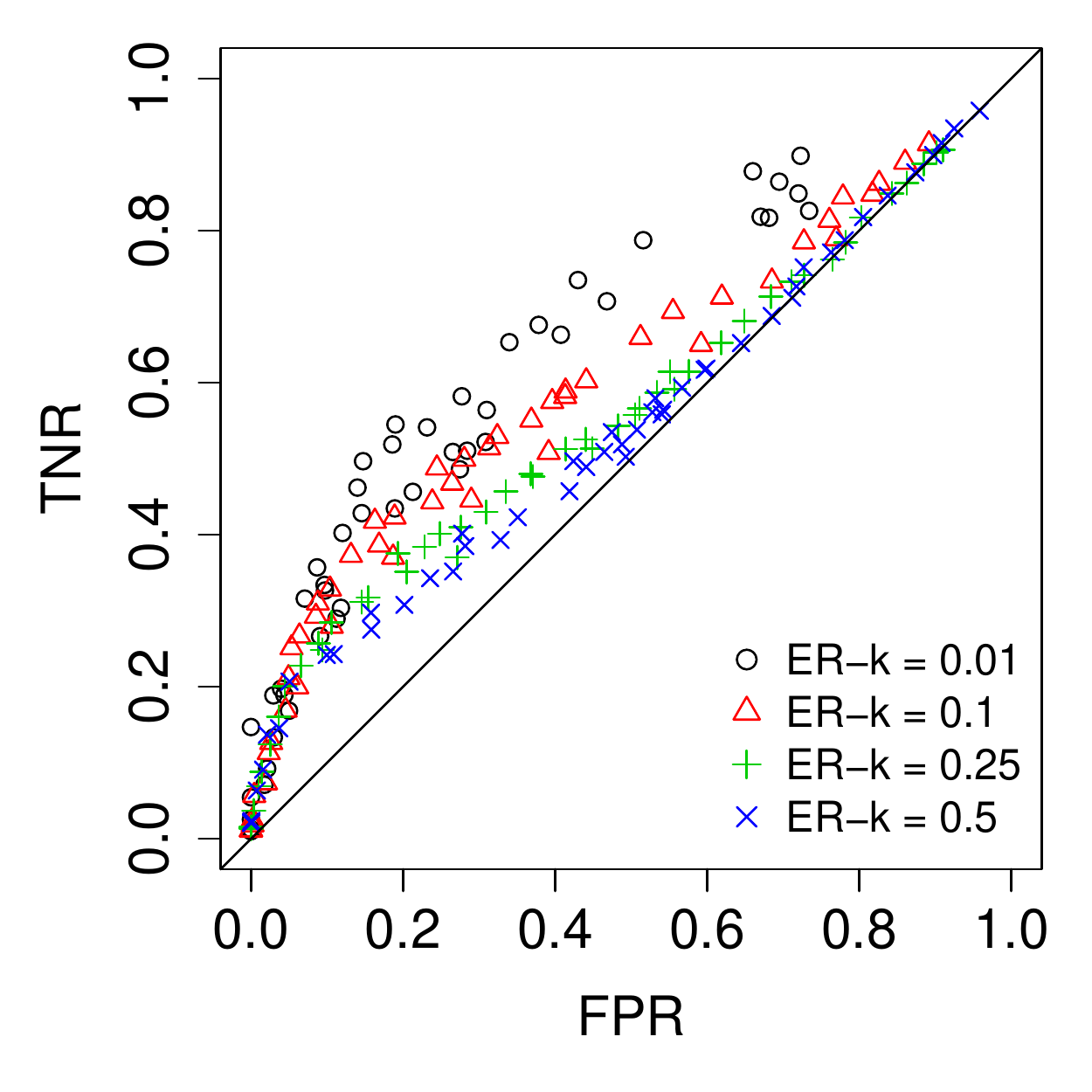}
\includegraphics[width=.49\linewidth]{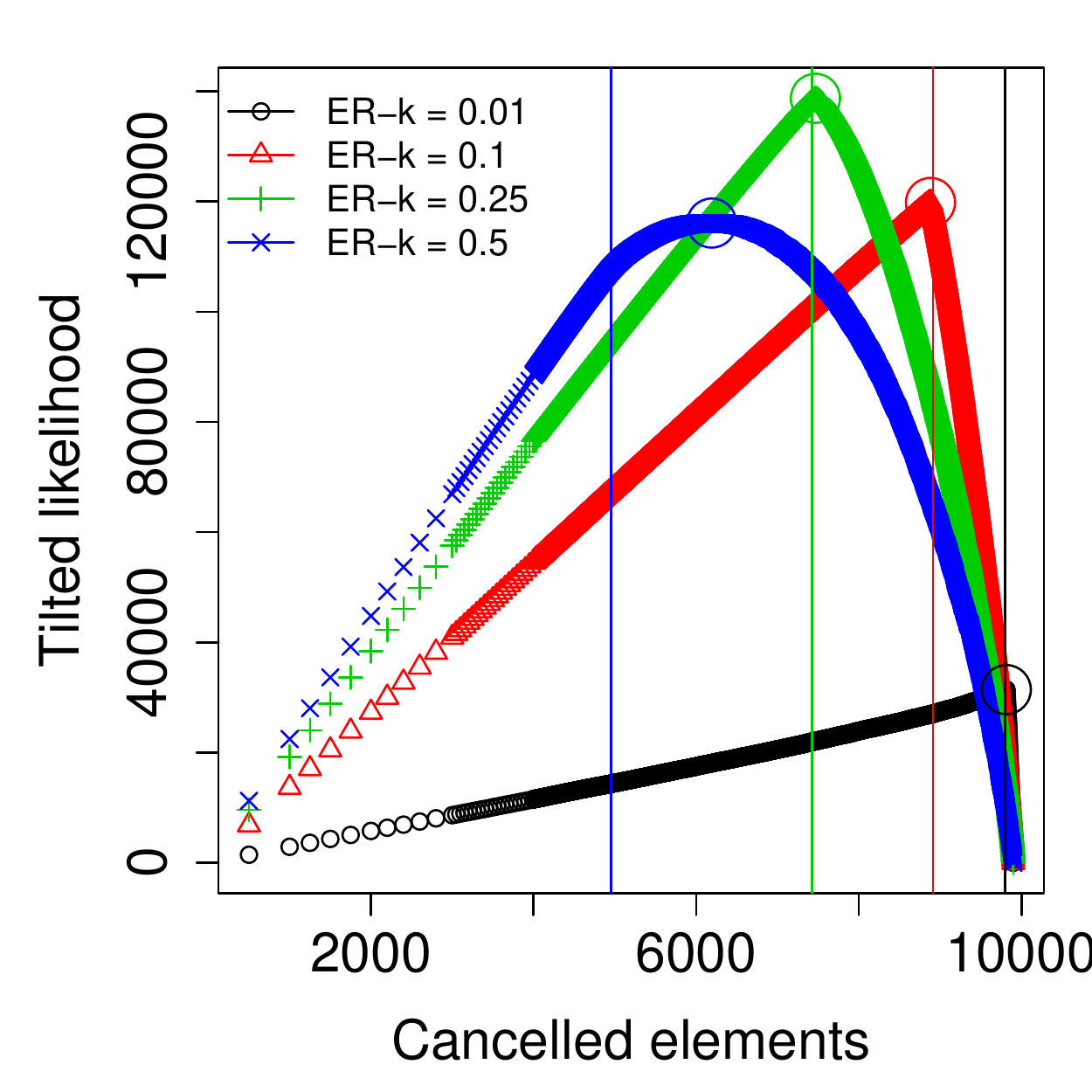}
\includegraphics[width=.49\linewidth]{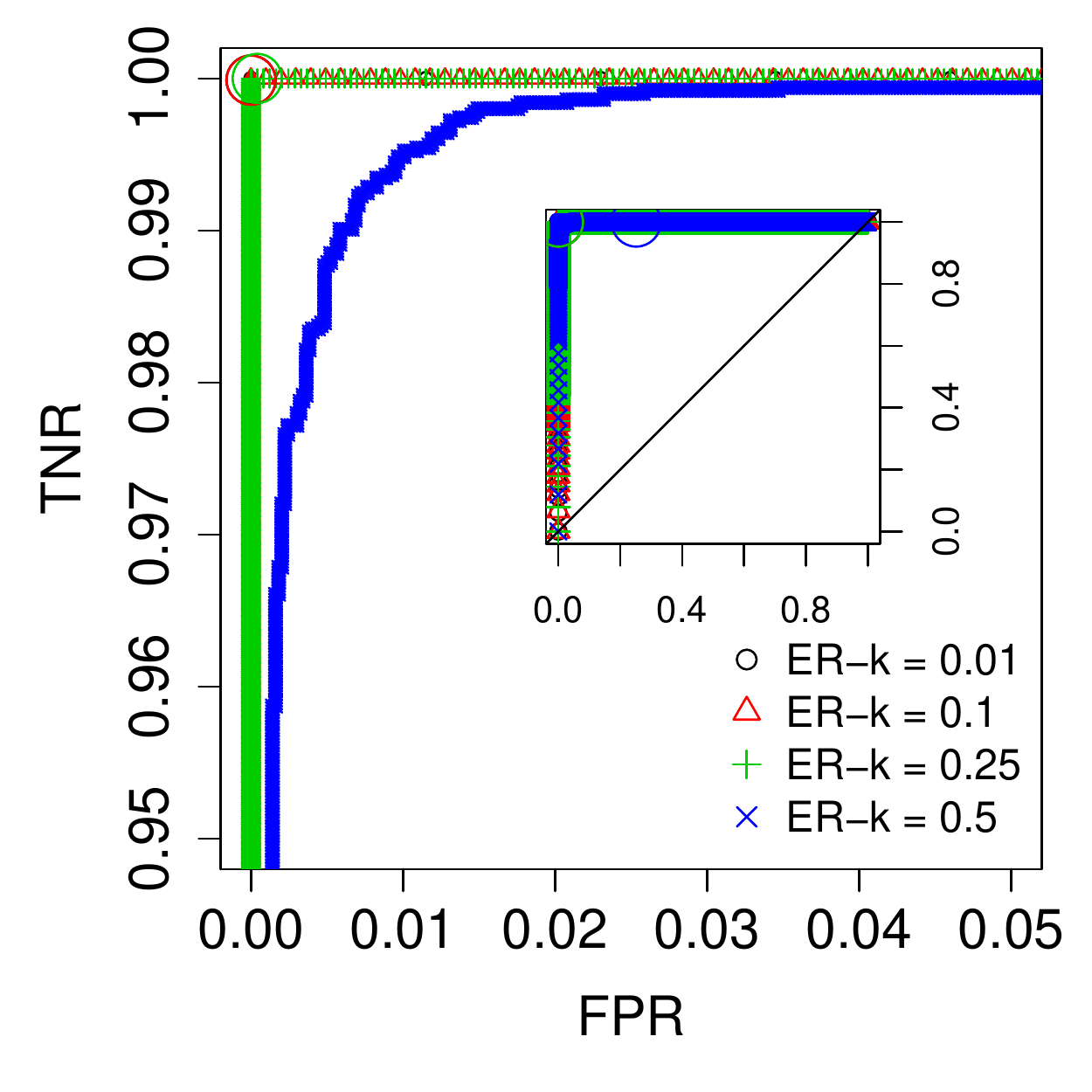}
\caption{(top) Results from the LASSO with 80\% observations: (left) RMSE on couplings as a function of the LASSO parameter; (right) ROC curves. (bottom) Results from the decimation procedure with 80\% observations: (left) Tilted likelihood evolution through the decimation process, vertical lines show the correct number of null elements; (right) ROC curves through the decimation process with different network densities. The circle identifies the point at which the Tilted Likelihood is maximized.} \label{fig:test3}
\end{figure}

\begin{figure}
    \centering
    \includegraphics[width=.49\linewidth]{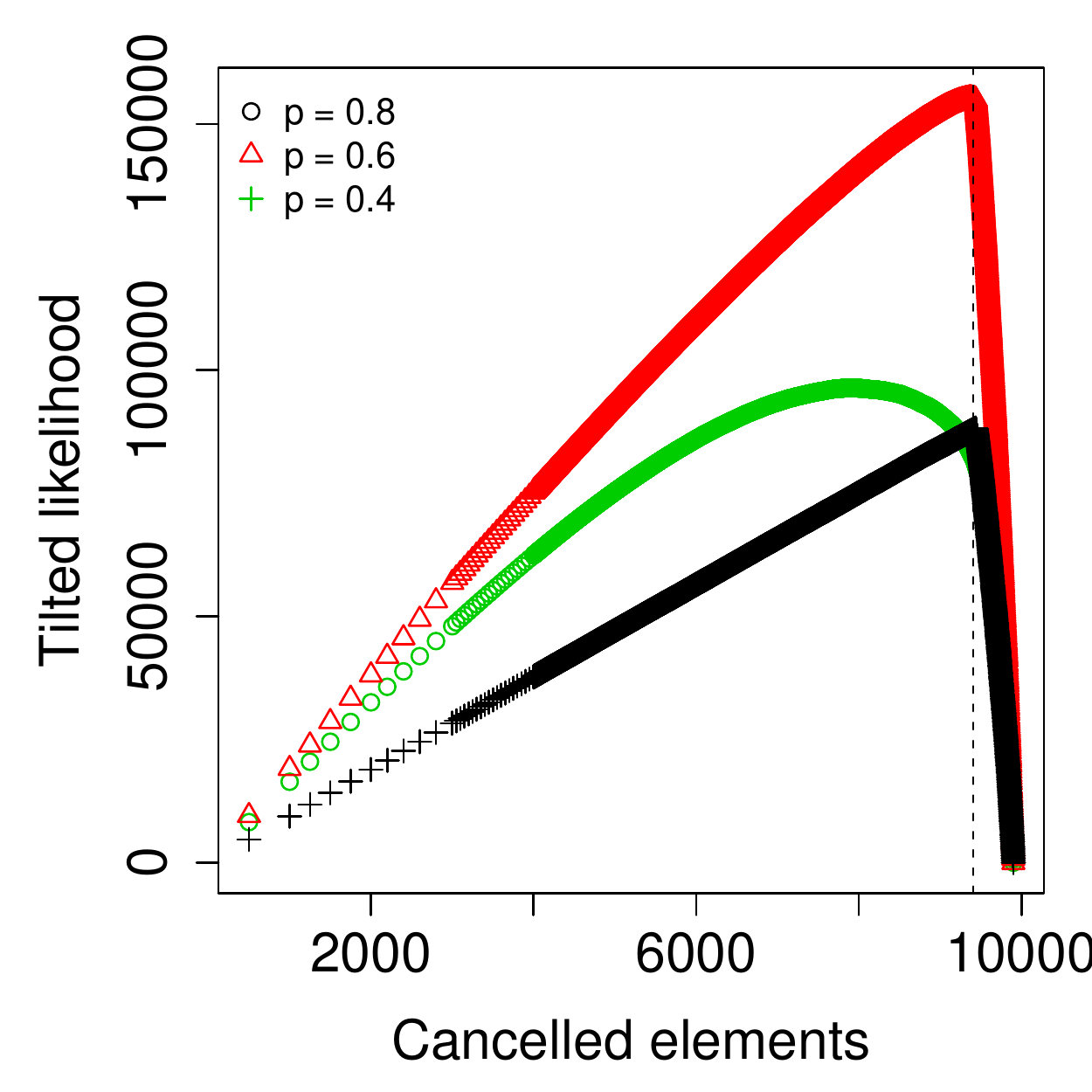}
    \includegraphics[width=.49\linewidth]{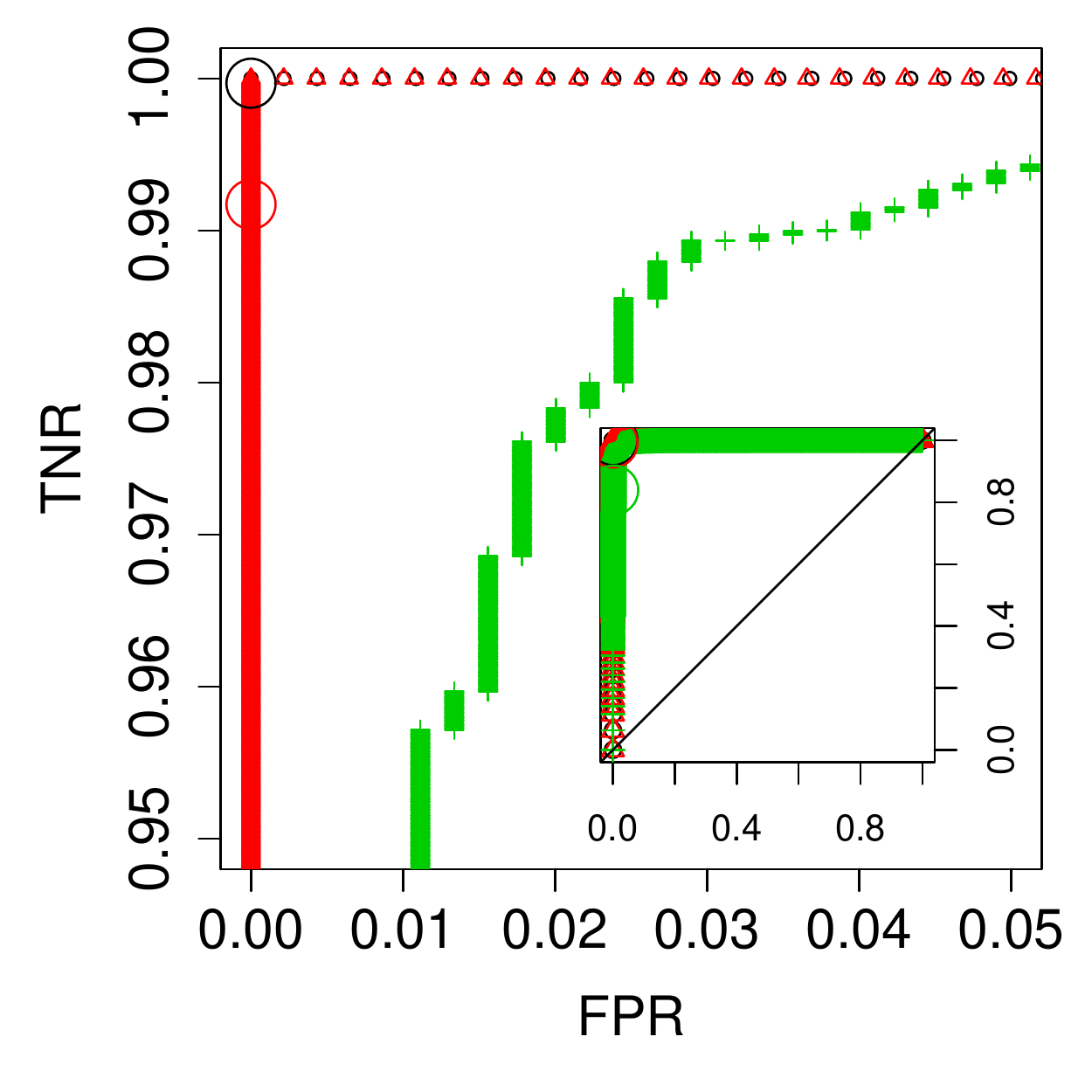}
    \includegraphics[width=.49\linewidth]{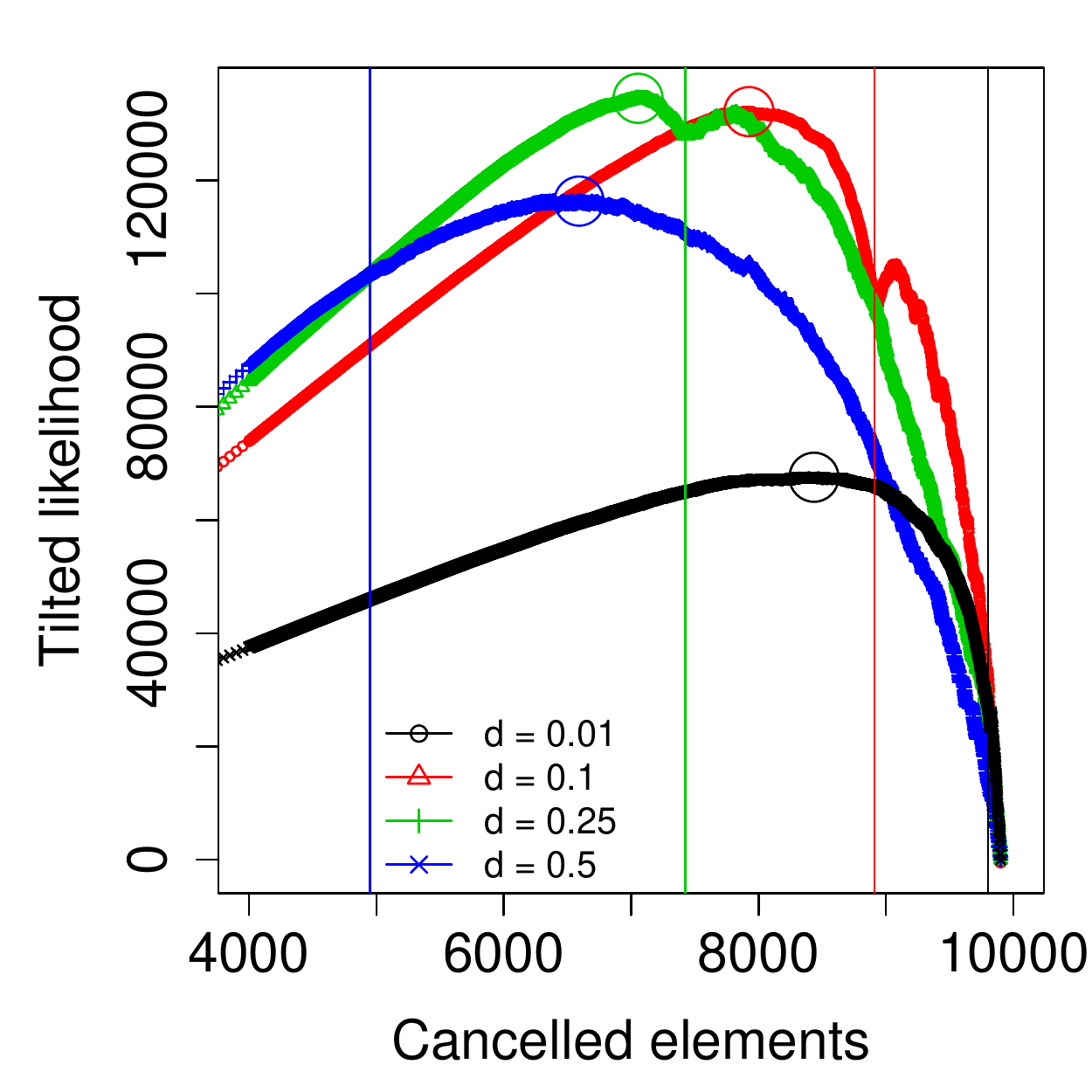}
    \includegraphics[width=.49\linewidth]{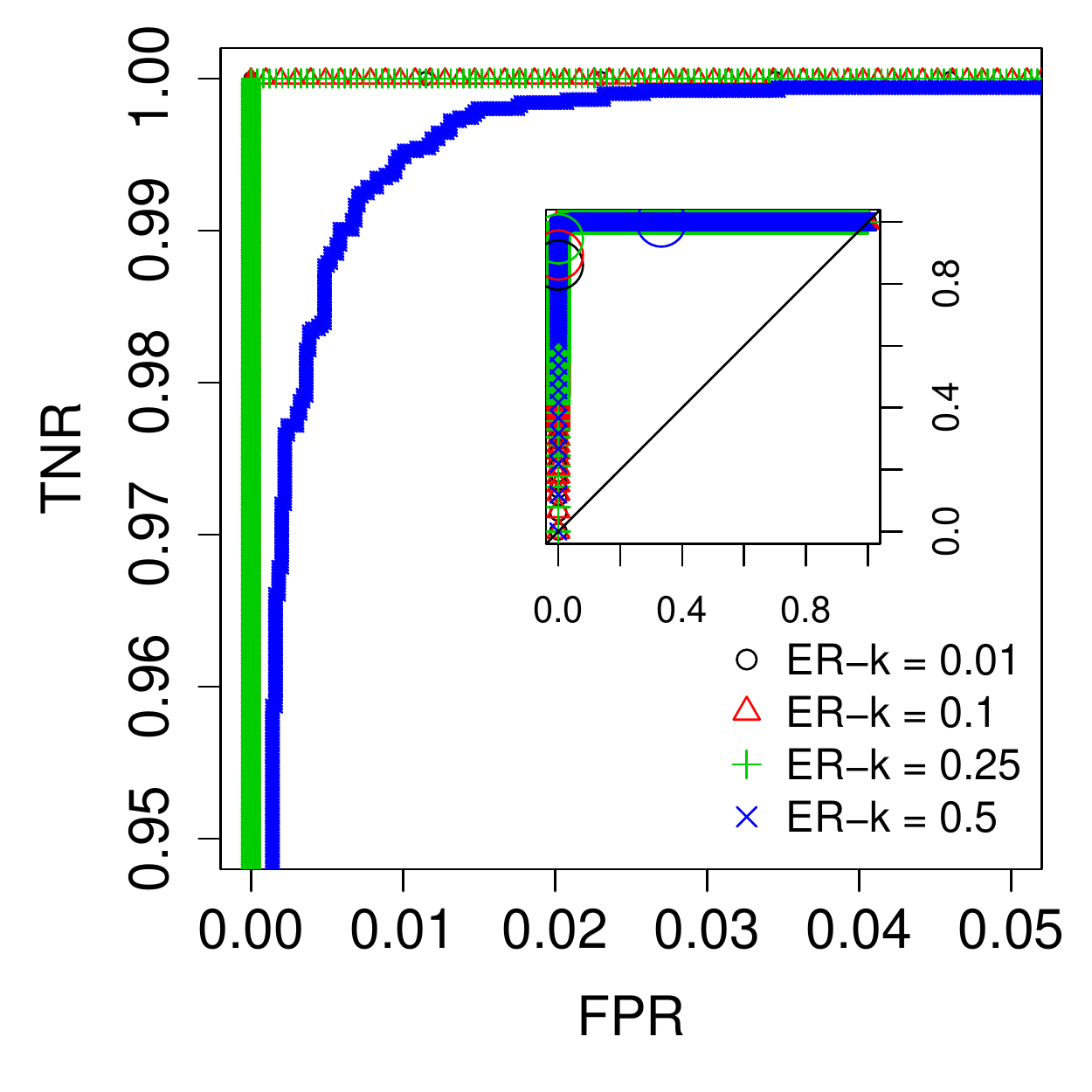}
    \caption{(top) Results from the decimation procedure with 80\%, 60\% and 40\% observations available and a network density of $0.05$: (left) Tilted Likelihood evolution through the decimation, vertical line shows the correct number of null elements; (right) ROC curves through the decimation process with different observation densities. (bottom) Results from the decimation introducing local fields $h$: (left) Tilted likelihood, vertical lines show the correct number of null elements; (right) ROC curves. The introduction of local fields makes the tilted likelihood non-convex and seriously affects the performance.}
    \label{fig:test3b}
\end{figure}

We test the algorithm performance on some more realistic network structure than the fully connected one. It is indeed known that real networks, and particularly social networks, are typically sparse and thus network models have to implement some pruning mechanism permitting to discriminate between noise, spurious correlations and actual causal relations. We generate our data simulating the Kinetic Ising model on one of the simplest random network models, the Erd\H{o}s-R\'{e}nyi model, with edges that have weights $J_{ij}$ normally distributed with variance $1/N$, $N=100$ and $T=10000$ and with a probability of observing the variables of $p \in \lbrace 0.8, 0.6, 0.4 \rbrace$. One then needs to adjust the algorithm to give sparse solutions, as the mean field approximation will tend to return fully connected $J$ matrices. The adjustments we make are the LASSO regularization and the decimation procedure of Decelle et al. \cite{decelle2015inference}. The first is the well known $\ell_1$ norm regularization of the objective function, which projects the maximum likelihood fully connected solution on a symplex of dimensions determined by a free parameter $\lambda$ (which has to be validated out of sample).\\
The second is a recently proposed technique that selects parameters starting to decimate them from the least significant ones and repeating the process until a so-called Tilted log-Likelihood function shows a discontinuity in the first derivative.\\
To briefly describe the procedure, call $\mathcal{L}_{max}$ the value of the log-likelihood provided by the maximum likelihood algorithm without any constraint and then call $x$ the fraction of parameters $J_{ij}$ that are being set to $0$. Finally call $\mathcal{L}(x)$ the log-likelihood of the model with the fraction $x$ of decimated parameters and $\mathcal{L}_1$ the log-likelihood of a model with no couplings that is, in case $h_i =0 \, \forall \, i$, $\mathcal{L}_1 = - \sum_t M(t) \log 2$. The Tilted log-Likelihood takes the form

\begin{equation*}
\mathcal{L}^{tilted}(x) = \mathcal{L}(x) - \left((1-x) \mathcal{L}_{max} + x \mathcal{L}_1 \right)
\end{equation*}
that is, the difference between a convex combination of the original log-likelihood with the log-likelihood of a system with no parameters and the log-likelihood of the decimated model. This function is strictly positive and is $0$ only for $x=0,1$, since $\mathcal{L}(0)=\mathcal{L}_{max}$ and $\mathcal{L}(1) = \mathcal{L}_1$, thus there has to be a maximum. The decimation process thus consists in gradually increasing the fraction of pruned parameters $x$ until the maximum of the Tilted log-Likelihood is found, giving the optimal set of parameters of the model.\\
We show in Figure \ref{fig:test3} and \ref{fig:test3b} the results of the test. We observe how the ROC curves seem to lean strongly in favor of the decimation approach, which tends to score perfectly on the False Positives Ratio (FPR) - True Negatives Ratio (TNR) plane. However the maximum of the Tilted Likelihood does not always correspond to the optimal score in the ROC diagram, both in the case of a non-sparse network and when the data has a large number of missing values. While the former case is not particularly interesting in that a dense network model fitted on real data would be prone to overfitting and of disputable use, the latter is much more of a concern, albeit the process is still surprisingly efficient even when data is extremely sparse.\\

Even if the decimation procedure is consistently outperforming the LASSO, there is reason to still hold the $\ell_1$ regularization as a viable option. Indeed when one introduces local fields $h$ of non-negligible entity, the decimation procedure is not anymore reliable in that the Tilted Likelihood becomes non-convex as shown in Figure \ref{fig:test3b} and the maximum is not in the correct position. This is due to the underestimation of the $h$ parameters during the log-likelihood maximization of the fully connected model, where part of the role of the local fields is absorbed in couplings that should be pruned. However these couplings are still relevant to the model since they compensate for the underestimated $h$ parameters, giving the Tilted likelihood a non-convex form and shifting its maximum towards a more dense network model. This situation does not occur with the LASSO regularization as the pruning is performed at the same time as the maximization, giving the LASSO the advantage of a much more reliable fit of the local fields albeit with an overall worse performance in the inference of the nonzero couplings.

\subsection{Test 4: Impact of asymmetricity assumption}
\begin{figure}
\includegraphics[width=1\linewidth]{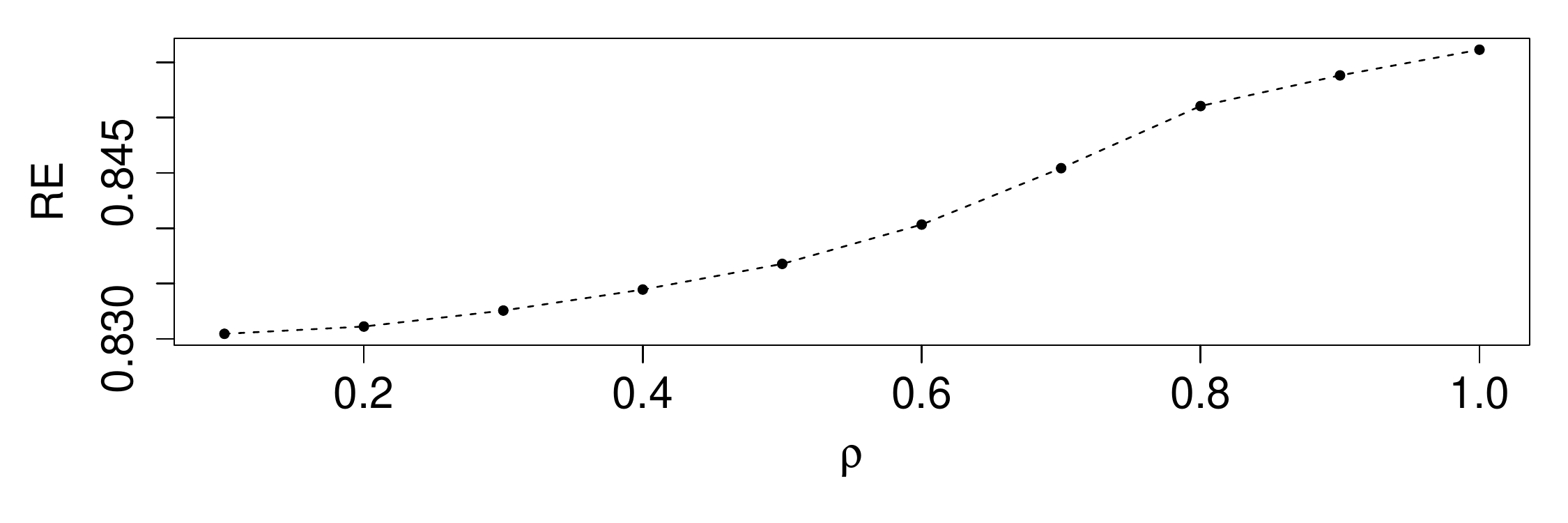}
\includegraphics[width=1\linewidth]{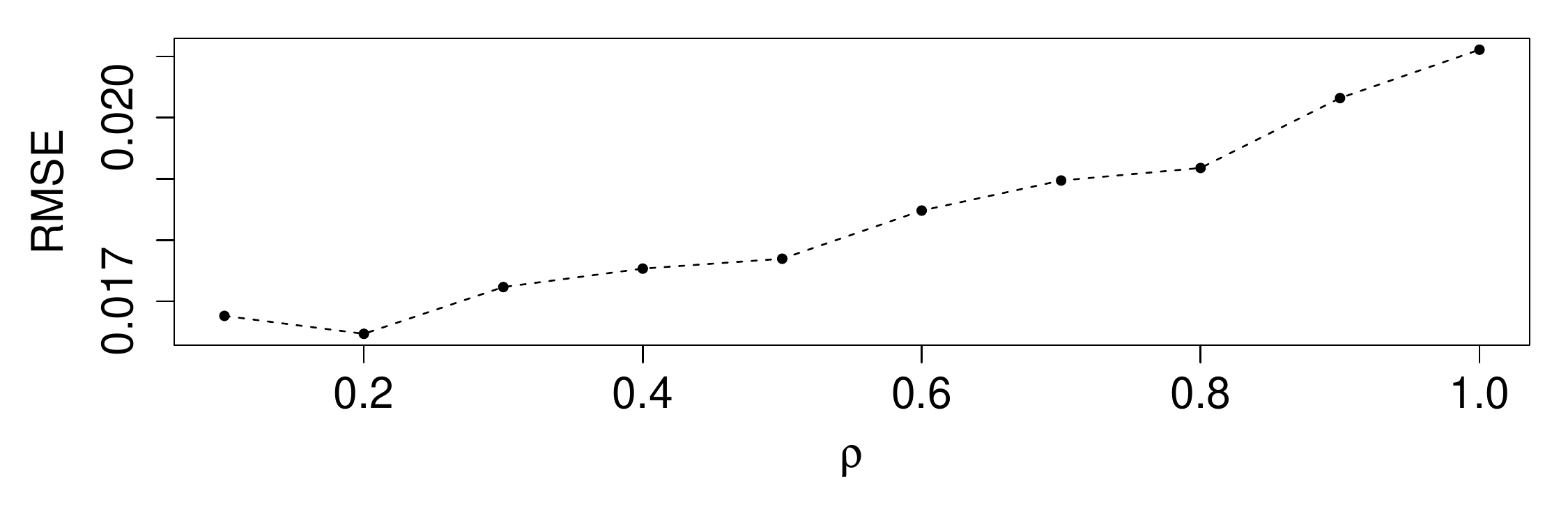}
\caption{(top) Reconstruction Efficiency varying the correlation between symmetric elements of $J$; (bottom) RMSE on the couplings.}\label{fig:rho}
\end{figure}

\begin{figure}
\includegraphics[width=1\linewidth]{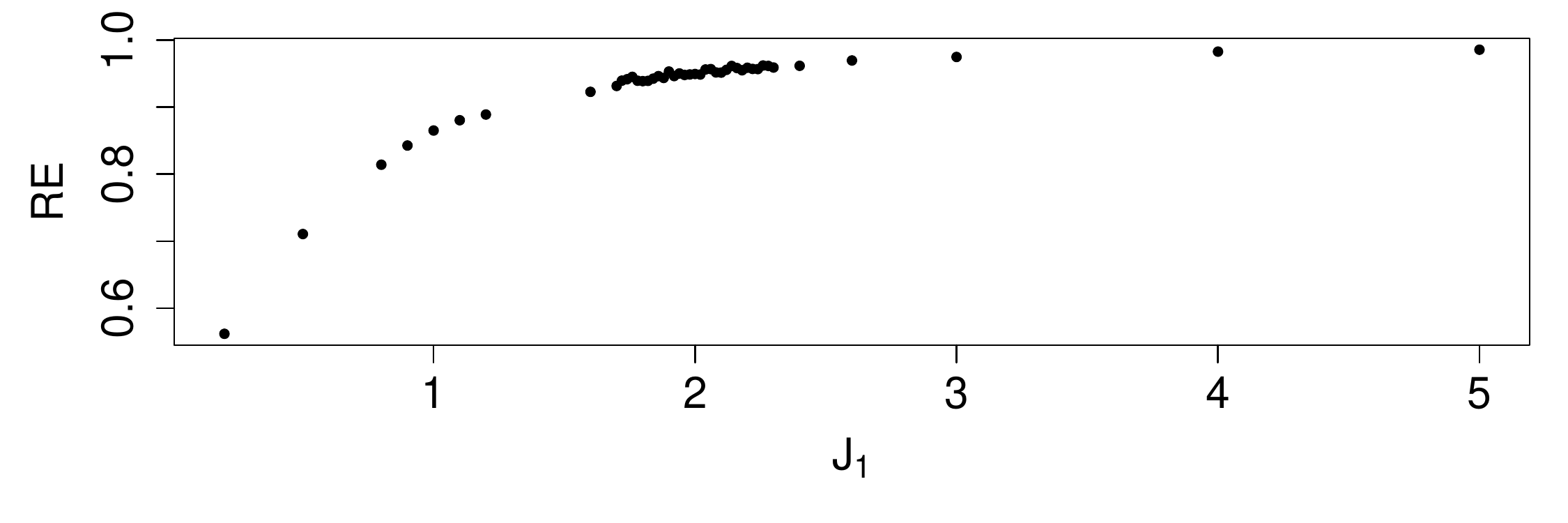}
\includegraphics[width=1\linewidth]{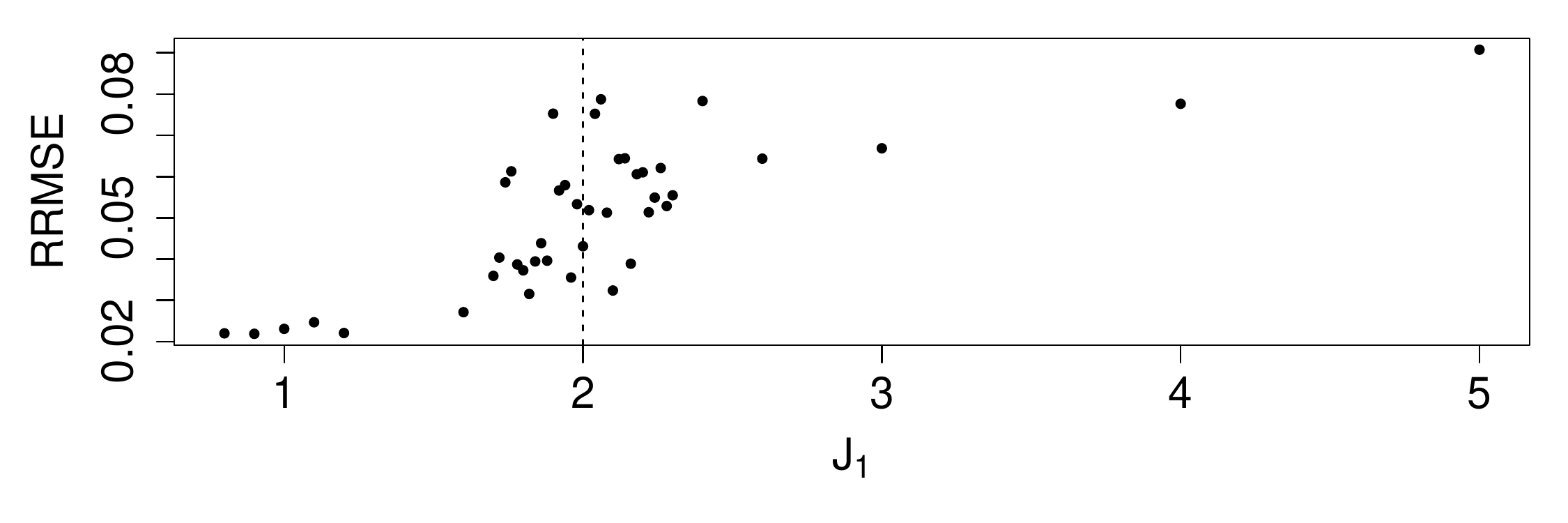}
\caption{(top) Reconstruction Efficiency as a function of $J_1$ in the SK model; (bottom) Rescaled RMSE on couplings as a function of $J_1$.}\label{fig:SK}
\end{figure}

Another assumption we made to perform the calculations in Equation \ref{eq:approx} was that the $J_{ij}$ are iid Gaussian random variables. In the case of social networks and trade networks reciprocity, that is the correlation between $J_{ij}$ and $J_{ji}$, is often found to be much higher than what would be expected in an iid setting \cite{squartini2013reciprocity}. We ask ourselves how impactful is this assumption on the outcome of the inference and we test the algorithm on data generated from a model with $N=100$, $T=10000$, $p_i=p=0.8$, $J_1=1$ and such that $\mathrm{Cor}(J_{ij}, J_{ji}) = \rho, \; i \neq j$. We show the results for this series of tests in Figure \ref{fig:rho}. What we find is that the $\rho$ parameter barely affects the performance and even makes it easier to infer the hidden variables, albeit marginally. Indeed we only used the assumption to approximate the determinant of the Hessian in the second order correction to the saddle-point solution, and letting the couplings not be reciprocally independent should affect the approximation slightly by having some elements of $J^2$ that vanish slower than others in the sums. It is possible that having a large enough $N$ facilitates the inference then, since the amount of those slowly vanishing terms grows with $N$ while the number of entries of $J$ grows with $N^2$.\\
We then turn our attention to the extreme case of $\rho=1$, corresponding to the well known Sherrington-Kirkpatrick (SK) model \cite{kirkpatrick1978infinite}, one of the first and most studied spin glass models in the literature.
The SK model has the peculiarity of undergoing a phase transition at $J_1=2$ in our notation for the Hamiltonian (since we have not included a factor $1/2$ to remove double counting), where for $J_1>2$ the spin glass phase arises and multiple equilibrium states appear such that the model is not easy to infer anymore. It is thus interesting to see whether this affects the inference from dynamical configurations and how the identifiability transition is reached. We perform the experiment of varying $J_1$ in this framework and show the results in Fig. \ref{fig:SK}. We find the expected increase in rescaled error (that is, $\mathrm{RMSE}/J_1$) marking the transition, surrounded by a finite-size scaling noisy region, while the reconstruction efficiency of the configurations remains very good. This fits in the narrative of the phase transition of the SK model, since in the spin glass phase an equilibrium configuration of the model can be generated by multiple - and in principle undistinguishable - choices of parameters which we indeed struggle to identify with our methodology.

\subsection{Test 5: sample size and convergence}

\begin{figure}
\includegraphics[width=1\linewidth]{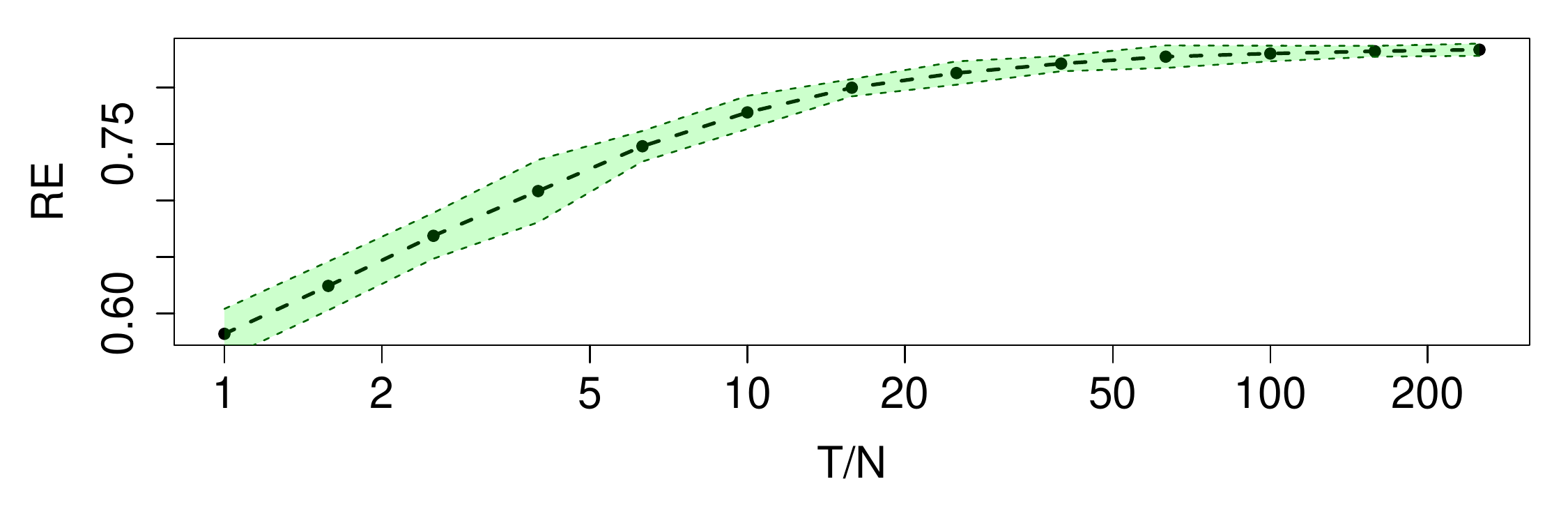}
\includegraphics[width=1\linewidth]{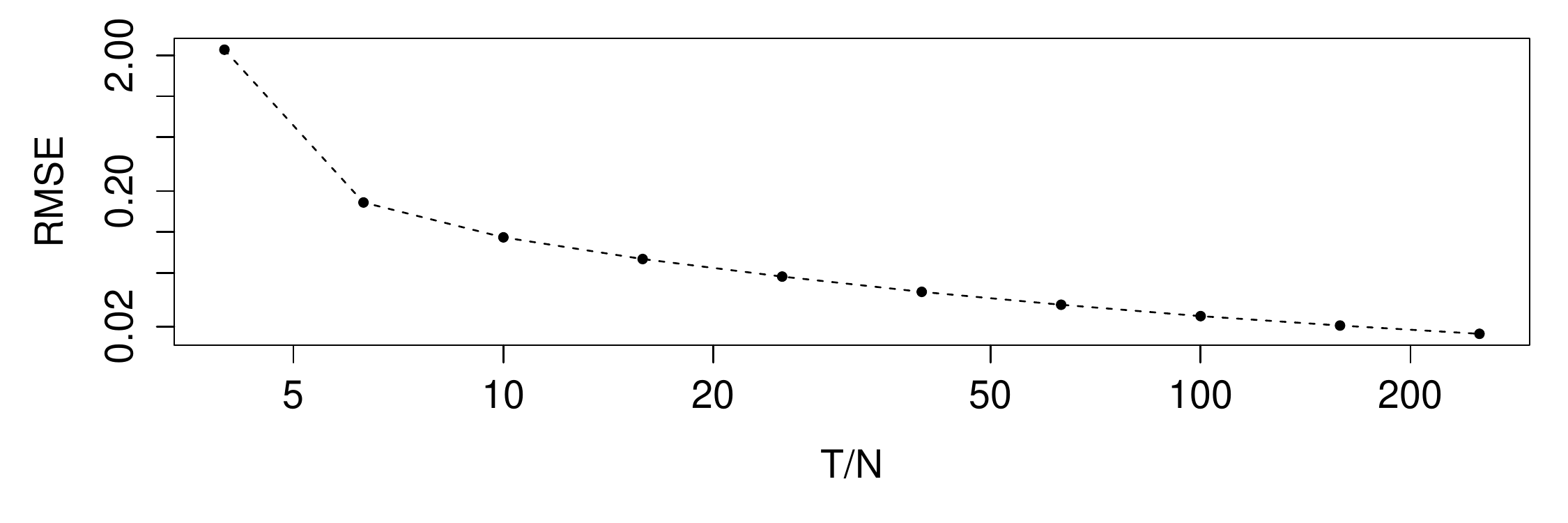}
\caption{(top) Reconstruction Efficiency as a function of the T/N ratio; (bottom) RMSE as a function of the T/N ratio. Area in green is 1 standard deviation from the mean over 30 repetitions.}\label{fig:tdep}
\end{figure}

We finally devolve our attention to the convergence properties of our estimator and how they are affected by finite sample sizes. The relevant parameter to be varied is the ratio between the length of the time series $T$ and the number of units that are modelled, $N$. We run simulations with $N=100$, $J_1 = 1$, $p_i = p = 0.8$ and varying $T$ between $100$ and $25000$, and report the results in Figure \ref{fig:tdep}. It can be seen that the RMSE on $J_{ij}$ diminishes, after $T/N=20$, with what might look like a power law behaviour with exponent close to $0.5$, although we do not provide an exact law for the convergence. The RMSE is below $5\%$ of $J_1$ when $T/N$ is larger than $20$ and is steadily converging towards $0$. Regarding the reconstruction efficiency we see that it saturates quickly towards $90\%$ and then it keeps increasing towards $100\%$. This evidence is an heuristic proof that the estimator is converging and is important to estimate how reliable a result might be given the $T/N$ ratio of the data. Although a more rigorous law would be much more appealing for the task, it would require being able to write the posterior of $J,\sigma$ given $s$, which to the best of our knowledge is not a feasible calculation in this setting.

\subsection{Additional parameters: exogenous drivers}

The model can be easily extended to a version in which an exogenous driver (or multiple ones), observed at all times, affects the dynamics of the variables. In a financial setting the first external driver would be given by the log-returns $r_t$ and the associated parameter would be the typical reaction of a trader to price changes, typically categorized between contrarians and chartists whether they go ``against" the flow (i.e. sell when the price rises and viceversa) or follow the trend. In the model, this is introduced by adding a set of linear parameters $\beta$ in the local fields that couple the variables to the driver
\begin{equation*}
g_k(t) = \sum_l y_l (t) + h_k + \beta_k r_t
\end{equation*}

The introduction of the parameter does not complicate the inference process at all and is particularly important if one wants to use the model to describe and possibly forecast order flows in financial markets. We omit the results for this section for the sake of space and because no significant dependency on the size of the $\beta_k$ parameters is found for our performance metrics.


\section{Conclusions}

In this article we develop a methodology to perform inference of Kinetic Ising Models on datasets with missing observations. We successfully adapt a known approximation from the Mean Field literature to the presence of missing values in the sample and devise several performance tests to characterize the algorithm and show its potential. We also propose a recursive methodology, R-EM, that gradually reconstructs the dataset with inferred quantities and tries to refine the inference, and show its efficacy on synthetic data.\\
The main results are that it is indeed possible to infer Kinetic Ising Models from incomplete datasets and that our procedure is resilient to noise, heterogeneity in the nature of data and in the frequency of missing values, and overall quantity of missing data. We make the algorithm ready for real-world applications by implementing pruning techniques in the form of LASSO and decimation, and give a brief overview of what we think are the better uses for each.\\
The methodology lends itself to applications on many diverse datasets, but our main focus for future research will be on opinion spreading in financial markets where transactions occur at high frequency, such as the FX or the cryptocurrency markets. We indeed envision our algorithm can identify significant structures of lagged correlations between traders, that in turn can be mapped to a network of lead-lag relations. Such a network would be particularly useful to get a quantitative picture of how possible speculative or irrational price movements can occur due to voluntary or involuntary coordination between traders and to devise appropriate strategies to counteract them.
\vfill
\section*{Acknowledgements}
The authors are grateful to prof. Matteo Marsili and to the participants of the 2018 Spring College on the Physics of Complex Systems (Trieste) for insightful comments and discussions. DT acknowledges GNFM-Indam and SNS for financial support of the project SNS18\_A\_TANTARI.

\appendix

\section{Zero-order saddle-point approximation}
We start from Eq.8 in the main text, where we have introduced the Dirac delta function to obtain a functional form of $\mathcal{L}$ for which the trace can be calculated. The result is the functional $\Phi$ of Eq.8, which once the trace is done reads
\begin{align*}
\Phi &= \sum_t \Bigg[ \sum_{i} \left[ s_i g_i - \log 2\cosh (g_i) \right] - \sum_{a} \log 2 \cosh (g_a) + \\ 
&+ \sum_{i} i \hat{g}_i \left[ g_i - \sum_{j} J^{oo}_{ij} s_j^- - h_i \right] + \\
&+ \sum_{a} i \hat{g}_a \left[g_a - \sum_{j} J^{ho}_{aj} s_j^- - h_a \right] + \\
&+ \sum_{a} \log 2 \cosh \left[g_a^- - \sum_{i}i \hat{g}_i J^{oh}_{ia} - \sum_{b} i \hat{g}_b J^{hh}_{ba} + \psi_a^- \right] \Bigg]
\end{align*}
This is the function to be extremized to find the saddle-point around which the integral is to be computed. Setting $\nabla_{\mathcal{G}} \Phi = 0$ gives
\begin{align*}
&g_i^0 = h_i + \sideset{}{^-}\sum_{j} J^{oo}_{ij}s_j^- + \sideset{}{^-}\sum_{a} J^{oh}_{ia} m_a^-  \\
&g_a^0 = h_a + \sideset{}{^-}\sum_{j} J^{ho}_{aj}s_j^- + \sideset{}{^-}\sum_{b} J^{hh}_{ab} m_a^- \\
&i\hat{g}_i^0 = \tanh (g_i) - s_i \\
&i\hat{g}_a^0 = \tanh (g_a) - m_a \\
\end{align*}
which, substituted in $\Phi$, give the zero-order solution to the saddle-point integral. The other ingredient is the vector of magnetizations $m$ which, as stated in the main text, is obtained exploiting the property of $\mathcal{L}$ being the moment generating functional for $\sigma$. Thus we find
\begin{equation*}
\lim_{\psi_a \rightarrow 0}\frac{\partial \mathcal{L}}{\partial \psi_a} = m_a  = \tanh \left[ g_a^0 - \sideset{}{^\prime}\sum_{i} i \hat{g}_i^{0\prime} J^{oh}_{ia} - \sideset{}{^\prime}\sum_{b} i \hat{g}_b^{0\prime} J^{hh}_{ba} \right]
\end{equation*}



\section{Second order saddle-point approximation}
The second order approximation requires the calculation of the determinant of the Hessian of the log-likelihood, $\nabla^2_{\mathcal{G}} \mathcal{L}$, taken at the saddle point coordinates. This is a forbidding task to tackle numerically, since the matrix has $(4NT)^2$ elements, but with a few algebraic manipulations the computations become feasible.
\begin{widetext}
The Hessian matrix elements can be summarized in the following sub-matrices $A^{tt'}, ..., G^{tt'}$, given by
\begin{align*}
&\frac{\partial^2 \Phi}{\partial g_i(t) \partial g_j(t')} = A_{ij}^{tt'} = - \delta_{ij} \delta_{tt'} (1 - \tanh^2 [g_i^0(t)])\\
&\frac{\partial^2 \Phi}{\partial \hat{g}_i(t) \partial \hat{g}_j(t')} = B_{ij}^{tt'} = - \delta_{tt'} \sideset{}{^-}\sum_{a} J_{ia}^{oh}(t) J_{ja}^{oh}(t) [1 - \mu_a^2 (t-1)]  \\
&\frac{\partial^2 \Phi}{\partial g_a(t) \partial g_b(t')} = C_{ab}^{tt'} = -\delta_{ab}\delta_{tt'}\left[\mu_a^2(t) - \tanh^2[g_a^0(t)] \right] \\
&\frac{\partial^2 \Phi}{\partial \hat{g}_a(t) \partial \hat{g}_b(t')} = D_{ab}^{tt'} = - \delta_{tt'} \sideset{}{^-}\sum_{c} J_{ac}^{hh}(t) J_{bc}^{hh} (t) \left[ 1 - \mu_c^2(t-1) \right] \\
&\frac{\partial^2 \Phi}{\partial \hat{g}_i(t) \partial \hat{g}_b(t')} = E_{ib}^{tt'} = - \delta_{tt'} \sideset{}{^-}\sum_{a} J_{ia}^{oh}(t) J_{ba}^{hh}(t) \left[ 1 - \mu_a^2(t-1) \right] \\
&\frac{\partial^2 \Phi}{\partial \hat{g}_i(t) \partial g_b(t')} = F_{ib}^{tt'} = -i \delta_{t-1, t'} J_{ib}^{oh}(t) \left[ 1 - \mu_b^2(t-1) \right] \\
&\frac{\partial^2 \Phi}{\partial g_a(t) \partial \hat{g}_b(t')} = \delta_{ab} \delta_{tt'} + G_{ab}^{tt'} = \delta_{ab} \delta_{tt'} - i \delta_{t+1, t'} J_{ba}^{hh}(t+1) \left[ 1 - \mu_a^2 (t) \right] \\
&\frac{\partial^2 \Phi}{\partial g_i(t) \partial \hat{g}_j (t')} = \delta_{ij} \delta_{tt'} \\
&\frac{\partial^2 \Phi}{\partial g_i(t) \partial g_b(t')} = \frac{\partial^2 \Phi}{\partial g_i(t) \partial \hat{g}_b(t')} = 0 \qquad \forall \: t, t', i, b
\end{align*}

and in matrix form it has the following almost block-diagonal form (we show the sub-matrix for times $t,t+1$)
\[
\left[
\begin{array}{cccc|cccc}
A^{tt} & i \mathbb{I} & 0 & 0 & 0 & 0 & 0 & 0  \\
i \mathbb{I} & B^{tt} & 0 & E^{tt} & 0 & 0 & 0 & 0 \\
0 & 0 & C^{tt} & i \mathbb{I} & 0 & \left[ F^{t+1,t} \right]^T & 0 & G^{t,t+1} \\
0 & \left[ E^{tt} \right]^T & i \mathbb{I} & D^{tt} & 0 & 0 & 0 & 0 \\
\hline
0 & 0 & 0 & 0 & A^{t+1,t+1} & i \mathbb{I} & 0 & 0 \\
0 & 0 & F^{t+1,t} & 0 & i \mathbb{I} & B^{t+1, t+1} & 0 & E^{t+1,t+1} \\
0 & 0 & 0 & 0 & 0 & 0 & C^{t+1,t+1} & i \mathbb{I} \\
0 & 0 & \left[ G^{t,t+1} \right]^T & 0 & 0 & \left[E^{t+1,t+1} \right]^T & i \mathbb{I} & D^{t+1,t+1}
\end{array}
\right]
\]
\end{widetext}
It is thus clear that the determinant of this matrix, under the approximation in Eq.9 of the main text, is
\begin{equation*}
    \det\left[ \nabla^2_\mathcal{G} \mathcal{L} \right] \approx \prod_t (\det A^{tt} \det B^{tt} + \mathbb{I} ) (\det C^{tt} \det D^{tt} + \mathbb{I})
\end{equation*}
which leads to the form of the correction reported in the main text.\\
As mentioned in Eq. \ref{eq:corrselfcon} in the main text, introducing the Gaussian correction shifts the magnetizations by a quantity
\begin{align*}
l_a(t) = &\frac{\partial( \delta \mathcal{L})}{ \partial \psi_a(t)} =\\
= \mu_a (1-&\mu_a^2) \left[ \sideset{}{'}\sum_{i} \left[ \left(1 - \tanh^2(g_i^\prime) \right) \left[J^{oh \prime}_{ia} \right]^2 \right] \right] \\ 
+ \mu_a (1-&\mu_a^2) \Bigg[ \sideset{}{^-}\sum_{b} \left[ J^{hh}_{ab} \right]^2 (1-\mu_b^{- \, 2}) + \\
&+ \sideset{}{'}\sum_{b} \left(\mu_b^{\prime \, 2} - \tanh^2(g_b^\prime) \right) \left[ J^{hh \prime}_{ba} \right]^2 \Bigg]
\end{align*}

Thus we rewrite both $\Gamma_0$ and $\delta \mathcal{L}$ substituting $\mu_a(t)\vert_{\psi_a(t)=0} = m_a(t) - l_a(t)\vert_{\psi_a(t) = 0}$ in the functional and in the saddle-point solutions for $g$ and obtain

\begin{align*}
\Gamma_0[m] = \sum_t \Bigg[& \sideset{}{'}\sum_{i} \left[ s_i^\prime g_i^\prime - \log 2 \cosh(g_i^\prime) \right] + \\
 + &\sideset{}{'}\sum_{a} \left[ m_a^\prime g_a^\prime - \log 2 \cosh(g_a^\prime) \right] + \sum_{a} S[m_a] + \\
 - &\sideset{}{'}\sum_{i} \left[ s_i^\prime - \tanh (g_i^\prime) \right] \sum_{a} J^{oh \, \prime}_{ia} l_a + \\
 - &\sideset{}{'}\sum_{a} \left[m_a^\prime - \tanh(g_a^\prime) \right] \sum_{b} J^{hh \, \prime}_{ab} l_{b} + \\
 - &\sideset{}{'}\sum_{a} l_a^\prime \left[ g_a^\prime - \sum_{b} J^{hh \, \prime}_{ab} l_b \right] +\\
 + &\sum_{a} l_a \tanh^{-1} (m_a) \Bigg]
\end{align*}

Where in this last formula $g(t)$ have become the fields of Eq.4 in the main text with $m$ in place of $\sigma$.
Given this last expression it can be seen that, since $l_a(t)$ is already quadratic in $J$ and always multiplies an object of order one, all terms involving $l_a(t)$ are higher order and can be neglected in the current approximation. \\
Skipping to Eq.12 in the main text and adding the Gaussian correction to the $\Gamma_0$ functional we obtain the final form of the approximated log-likelihood to be maximized
\begin{align*}
\Gamma_1[m] &= \Gamma_0[m] +\\
&- \frac{1}{2} \sum_t \sideset{}{'}\sum_{i} \left[ \left(1 - \tanh^2(g_i^\prime)\right) \sum_{b} \left[J^{oh \, \prime}_{ib}\right]^2 (1 - m_b^2) \right] + \\
&- \frac{1}{2} \sum_t \sideset{}{'}\sum_{a} \left[ \left(m_a^{2 \, \prime} - \tanh^2(g_a^\prime)\right) \sum_{b} \left[ J^{hh \, \prime}_{ab} \right]^2 (1-m_b^2) \right]
\end{align*}

The final result are the formulas necessary to the EM-like algorithm, namely the log-likelihood gradient and the self-consistent relations for the magnetizations. The first takes the form

\begin{alignat*}{2}
\frac{\partial \Gamma_1}{\partial J_{kl}} & = \sum_t \Bigg[ \sideset{}{'}\sum_{i}  \left[ \frac{\partial g_i^\prime}{\partial J_{kl}} \left(s_i^\prime - \tanh(g_i^\prime)\right) \right] + \\
& + \sideset{}{'}\sum_{a} \left[ \frac{\partial g_a^\prime}{\partial J_{kl}} \left(m_a^\prime - \tanh(g_a^\prime)\right) \right] + \\
& + \sideset{}{'}\sum_{i} \left[  \frac{\tanh (g_i^\prime)}{\cosh^2 (g_i^\prime)} \frac{\partial g_i^\prime}{\partial J_{kl}} \sum_{bmn} G_{im}^\prime J_{mn}^2 F^T_{nb} (1-m_b^2) \right] +  \\
& + \sideset{}{'}\sum_{i} \left[  - \left(1-\tanh^2(g_i^\prime) \right)\sum_{b} G_{ik}^\prime J_{kl} F^T_{lb} (1-m_b^2) \right] + \\
& + \sideset{}{'}\sum_{a} \left[  \frac{\tanh (g_a^\prime)}{\cosh^2 (g_a^\prime)} \frac{\partial g_a^\prime}{\partial J_{kl}} \sum_{bmn} F_{am}^\prime J^2_{mn} F^T_{nb} (1-m_b^2) \right] + \\
& + \sideset{}{'}\sum_{a} \left[ - \left(m_a^{2 \, \prime} - \tanh^2(g_a^\prime) \right) \sum_{b} F_{ak}^\prime J_{kl} F^T_{lb}(1-m_b^2) \right] \Bigg] 
\end{alignat*}
\begin{widetext}
where the fields $g$ and their derivatives are given by
\begin{align*}
&g_i^\prime = \sum_{j} \sum_{kl} G_{ik}^\prime J_{kl} G^T_{lj} s_j + \sum_{b} \sum_{kl} G_{ik}^\prime J_{kl} F^T_{lb} m_b + h_i \\
&g_a^\prime = \sum_{j} \sum_{kl} F_{ak}^\prime J_{kl} G^T_{lj} s_j + \sum_{b} \sum_{kl} F_{ak}^\prime J_{kl} F^T_{lb} m_b + h_a \\
&\frac{\partial g_i^\prime}{\partial J_{kl}} = \sum_{j} G_{ik}^\prime G^T_{lj} s_j + \sum_{b} G_{ik}^\prime F^T_{lb} m_b \\
&\frac{\partial g_a^\prime}{\partial J_{kl}} = \sum_{j} F_{ak}^\prime G^T_{lj} s_j + \sum_{b} F_{ak}^\prime F^T_{lb} m_b 
\end{align*}

The self consistency equations for the magnetizations $m$ are then obtained by imposing $\partial\Gamma_1 / \partial m_a(t) = 0$, finding

\begin{alignat*}{2}
m_a = \tanh \Bigg[g_a &+ m_a \bigg[ &&\sideset{}{'}\sum_{i} \left(1-\tanh^2(g_i^\prime) \right) \sum_{kl} G_{ik}^\prime J_{kl}^2 F^T_{la} + \nonumber \\ 
& &&+ \sideset{}{'}\sum_{b} \left(m_b^{2 \, \prime} - \tanh^2 (g_b^\prime) \right) \sum_{kl} F_{bk}^\prime J_{kl}^2 F^T_{la} + \nonumber \\
& &&- \sideset{}{^-}\sum_{c}\sum_{kl} F_{ak} J_{kl}^2 F^{T \, -}_{lc} (1-m_c^{2 \, -}) \bigg] + \nonumber \\
&+ \sideset{}{'}\sum_{i} &&\left(s_i^\prime - \tanh (g_i^\prime) \right) \sum_{kl} G_{ik}^\prime J_{kl} F^T_{la} + \nonumber \\
&+ \sideset{}{'}\sum_{b} &&\left(m_b^\prime - \tanh (g_b^\prime) \right) \sum_{kl} F_{bk}^\prime J_{kl} F^T_{la} + \nonumber \\
&+ \sideset{}{'}\sum_i &&\frac{\tanh (g_i^\prime)}{\cosh^2 (g_i^\prime)} \sum_{oqb} G_{io}^\prime J_{oq} F^T_{qb} (1 - m_b^{2 \, \prime})  \sum_{kl} G_{ik}^\prime J_{kl} F^T_{la} + \nonumber \\
&+ \sideset{}{'}\sum_c &&\frac{\tanh (g_c^\prime)}{\cosh^2 (g_c^\prime)} \sum_{oqb} F_{co}^\prime J_{oq} F^T_{qb} \left(1-m_b^{2 \, \prime} \right) \sum_{kl} F_{ck}^\prime J_{kl} F^T_{la} \Bigg]  \label{selfcon::3}
\end{alignat*}
\end{widetext}


\bibliographystyle{unsrt}%
\bibliography{biblio.bib}%

\begin{thebibliography}{10}

\bibitem{bury2013market}
Thomas Bury.
\newblock Market structure explained by pairwise interactions.
\newblock {\em Physica A: Statistical Mechanics and its Applications},
  392(6):1375--1385, 2013.

\bibitem{Bouchaud2013}
Jean-Philippe Bouchaud.
\newblock Crises and collective socio-economic phenomena: Simple models and
  challenges.
\newblock {\em Journal of Statistical Physics}, 151(3):567--606, May 2013.

\bibitem{tanaka1977model}
Seiji Tanaka and Harold~A Scheraga.
\newblock Model of protein folding: incorporation of a one-dimensional
  short-range (ising) model into a three-dimensional model.
\newblock {\em Proceedings of the National Academy of Sciences},
  74(4):1320--1323, 1977.

\bibitem{cocco2017functional}
Simona Cocco, R{\'e}mi Monasson, Lorenzo Posani, and Gaia Tavoni.
\newblock Functional networks from inverse modeling of neural population
  activity.
\newblock {\em Current Opinion in Systems Biology}, 3:103--110, 2017.

\bibitem{kadirvelu2017inferring}
Balasundaram Kadirvelu, Yoshikatsu Hayashi, and Slawomir~J Nasuto.
\newblock Inferring structural connectivity using ising couplings in models of
  neuronal networks.
\newblock {\em Scientific reports}, 7(1):8156, 2017.

\bibitem{bornholdt2001expectation}
Stefan Bornholdt.
\newblock Expectation bubbles in a spin model of markets: Intermittency from
  frustration across scales.
\newblock {\em International Journal of Modern Physics C}, 12(05):667--674,
  2001.

\bibitem{ibuki2013statistical}
Takero Ibuki, Shunsuke Higano, Sei Suzuki, Jun-ichi Inoue, and Anirban
  Chakraborti.
\newblock Statistical inference of co-movements of stocks during a financial
  crisis.
\newblock In {\em Journal of Physics: Conference Series}, volume 473, page
  012008. IOP Publishing, 2013.

\bibitem{derrida1987exactly}
Bernard Derrida, Elizabeth Gardner, and Anne Zippelius.
\newblock An exactly solvable asymmetric neural network model.
\newblock {\em EPL (Europhysics Letters)}, 4(2):167, 1987.

\bibitem{crisanti1988dynamics}
A~Crisanti and Haim Sompolinsky.
\newblock Dynamics of spin systems with randomly asymmetric bonds: Ising spins
  and glauber dynamics.
\newblock {\em Physical Review A}, 37(12):4865, 1988.

\bibitem{capone2015inferring}
Cristiano Capone, Carla Filosa, Guido Gigante, Federico Ricci-Tersenghi, and
  Paolo Del~Giudice.
\newblock Inferring synaptic structure in presence of neural interaction time
  scales.
\newblock {\em PloS one}, 10(3):e0118412, 2015.

\bibitem{SornetteReview}
D.~Sornette.
\newblock {Physics and Financial Economics (1776-2014): Puzzles, Ising and
  Agent-Based models}.
\newblock Papers 1404.0243, arXiv.org, April 2014.

\bibitem{sakellariou2013inverse}
Jason Sakellariou.
\newblock {\em Inverse inference in the asymmetric ising model}.
\newblock PhD thesis, Universit{\'e} Paris Sud-Paris XI, 2013.

\bibitem{roudi2011dynamical}
Yasser Roudi and John Hertz.
\newblock Dynamical tap equations for non-equilibrium ising spin glasses.
\newblock {\em Journal of Statistical Mechanics: Theory and Experiment},
  2011(03):P03031, 2011.

\bibitem{dunn2013learning}
Benjamin Dunn and Yasser Roudi.
\newblock Learning and inference in a nonequilibrium ising model with hidden
  nodes.
\newblock {\em Physical Review E}, 87(2):022127, 2013.

\bibitem{ait2010high}
Yacine A{\"\i}t-Sahalia, Jianqing Fan, and Dacheng Xiu.
\newblock High-frequency covariance estimates with noisy and asynchronous
  financial data.
\newblock {\em Journal of the American Statistical Association},
  105(492):1504--1517, 2010.

\bibitem{buccheri2017score}
Giuseppe Buccheri, Giacomo Bormetti, Fulvio Corsi, and Fabrizio Lillo.
\newblock A score-driven conditional correlation model for noisy and
  asynchronous data: An application to high-frequency covariance dynamics.
\newblock {\em Available at SSRN: https://ssrn.com/abstract=2912438}, 2017.

\bibitem{Corsi2012}
Fulvio Corsi, Stefano Peluso, and Francesco Audrino.
\newblock Missing in asynchronicity: A kalman-em approach for multivariate
  realized covariance estimation.
\newblock {\em Journal of Applied Econometrics}, 30(3):377--397, 2015.

\bibitem{expmax1977}
A.~P. Dempster, N.~M. Laird, and D.~B. Rubin.
\newblock Maximum likelihood from incomplete data via the em algorithm.
\newblock {\em Journal of the Royal Statistical Society. Series B
  (Methodological)}, 39(1):1--38, 1977.

\bibitem{Nesterov2008}
Yu. Nesterov.
\newblock Accelerating the cubic regularization of newton's method on convex
  problems.
\newblock {\em Mathematical Programming}, 112(1):159--181, Mar 2008.

\bibitem{msr1973}
Paul~Cecil Martin, ED~Siggia, and HA~Rose.
\newblock Statistical dynamics of classical systems.
\newblock {\em Physical Review A}, 8(1):423, 1973.

\bibitem{tibshirani1996regression}
Robert Tibshirani.
\newblock Regression shrinkage and selection via the lasso.
\newblock {\em Journal of the Royal Statistical Society. Series B
  (Methodological)}, pages 267--288, 1996.

\bibitem{decelle2015inference}
Aur{\'e}lien Decelle and Pan Zhang.
\newblock Inference of the sparse kinetic ising model using the decimation
  method.
\newblock {\em Physical Review E}, 91(5):052136, 2015.

\bibitem{squartini2013reciprocity}
Tiziano Squartini, Francesco Picciolo, Franco Ruzzenenti, and Diego
  Garlaschelli.
\newblock Reciprocity of weighted networks.
\newblock {\em Scientific reports}, 3:2729, 2013.

\bibitem{kirkpatrick1978infinite}
Scott Kirkpatrick and David Sherrington.
\newblock Infinite-ranged models of spin-glasses.
\newblock {\em Physical Review B}, 17(11):4384, 1978.

\end{thebibliography}

\end{document}